\documentclass[aps,prc,nofootinbib,floatfix,twocolumn,superscriptaddess,showpacs,amsmath,amssymb]{revtex4-1}

\usepackage{dcolumn} 
\usepackage{graphicx} 
\usepackage{color}

\begin{document}
%
%

\title{Properties of trapped neutrons interacting with realistic nuclear Hamiltonians}

\author{Pieter Maris}
\affiliation{Department of Physics and Astronomy, Iowa State University, Ames, IA, 50011}

\author{James P. Vary}
\affiliation{Department of Physics and Astronomy, Iowa State University, Ames, IA, 50011}

\author{S. Gandolfi}
\affiliation{Theoretical Division,
Los Alamos National Laboratory, Los Alamos, NM 87545}

\author{J. Carlson}
\affiliation{Theoretical Division,
Los Alamos National Laboratory, Los Alamos, NM 87545}

\author{Steven C. Pieper}
\affiliation{Physics Division,
Argonne National Laboratory, Argonne, IL 60439}

\begin{abstract}
We calculate properties of neutron drops in external potentials using
both quantum Monte Carlo and no-core full configuration techniques.
The properties of the external wells are varied to examine different
density profiles.  We compare neutron drop results given by a
selection of nuclear Hamiltonians, including realistic two-body
interactions as well as several three-body forces.  We compute a range
of properties for the neutron drops: ground-state energies, spin-orbit
splittings, excitation energies, radial densities and rms radii.  We
compare the equations of state for neutron matter for several of these
Hamiltonians.  Our results can be used as benchmarks to test other
many-body techniques, and to constrain properties of energy-density
functionals.
\end{abstract}

\date{\today}
\pacs{21.30.-x, 21.60.-n, 21.60.De, 21.65.Cd}

\maketitle

\section{Introduction}

There are three major motivations for investigating pure neutron
systems with external fields (``neutron drops'') using {\it ab initio}
approaches.  First, neutron drops provide a very simple model of
neutron-rich nuclei in which the core is approximated as an external
well acting on valence
neutrons~\cite{Chang:2004a,Pieper:2005,Gandolfi:2006,Gandolfi:2008}.
Second, {\it ab initio} solutions for neutrons trapped by an external
potential can be used as data for calibrating model effective
Hamiltonians and Energy Density Functionals
(EDFs)~\cite{Bogner:2011kp,Gandolfi:2010za}, particularly for very
neutron-rich systems as occur in astrophysical environments like
neutron stars.  Third, these results may serve as useful benchmarks
for testing other many-body methods.

These motivations are further supported by the advent of new
experimental facilities to probe the extremes of neutron-rich nuclei,
to map out the neutron drip line and to inform models of nuclear
astrophysical processes~\cite{erler2012limits}.  Traditional energy
density functionals~\cite{bender2003self} are obtained by fitting
measured properties of stable and near-stable nuclei.  Extrapolations
using these traditional EDFs to regions of extreme isospin are
sensitive to their less controlled features that result in large
variations in the predictions.  Beyond these experimental vistas, we
desire control over properties of EDFs for low-density neutron
systems, since those properties are important for the processes in the
inner crust of neutron stars.  It is our long-term aim to provide {\it
  ab initio} calculations of trapped neutrons that address these
motivations.  We report here the results with currently available
approaches that serve as a basis for long-term efforts that will
employ improved microscopic Hamiltonians and many-body methods.

We adopt two nucleon-nucleon ($N\!N$) interactions which fit
scattering data and deuteron properties with high accuracy, namely the
local Argonne $v^\prime_8$ (AV8$^\prime$)~\cite{Pudliner:1997ck} and
the nonlocal JISP16~\cite{Shirokov07}.  As shown by accurate
calculations~\cite{Tucson,Urbana,Fadd-TNI,Pieper:2001,Pudliner:1997ck,Hayes:2003ni,Navratil:2007we,Maris:2011as,Roth:2011ar},
local $N\!N$ interactions are not sufficient to describe accurately
the properties of light nuclei.  Even the ground-state of the simplest
three-body problem, the triton, is significantly underbound.

Different models of three-nucleon interactions (TNIs) have been
proposed to build a non-relativistic Hamiltonian that reproduces
experimental results such as ground-state energies, density profiles,
and rms radii of light nuclei~\cite{Pieper:2001}.  TNIs from
meson-exchange theory are modeled through an operator structure with
parameters that are fit to experimental nuclear energies.  The Urbana
IX TNI (UIX) was fit to $^3$H and nuclear matter saturation, but it
typically underbinds heavier nuclei~\cite{Pudliner:1997ck}.  Other TNI
forms, namely Illinois forces, are fit to light
nuclei~\cite{Pieper:2001}.  The most recent is the Illinois-7
(IL7)~\cite{Pieper:2008} which reproduces nuclear energies up to
$A=12$ with an rms error of 600 keV.  However, the three-pion rings
included in IL7 give a strong overbinding of pure
neutron-matter~\cite{Sarsa:2003}.  In light of the different data
selected for tuning these TNIs it is useful to observe their
similarities and differences with neutron drops.

JISP16 is a phenomenological nonlocal $N\!N$-interaction written as a
finite matrix in a harmonic oscillator (HO) basis for each of the
$N\!N$ partial waves.  It is constructed to reproduce the available
$N\!N$ scattering data using the $J$-matrix inverse scattering
approach.  In addition, phase-equivalent transformations have been
used to modify its off-shell properties in order to achieve a good
description of selected states in light nuclei~\cite{Shirokov07}.  It
gives a good description of most narrow states in light nuclei up to
about $A=12$~\cite{Maris:2008ax,Maris:2009bx,Cockrell:2012vd}.
However it tends to overbind heavier $N=Z$ nuclei ($^{16}$O is
overbound by about 15\%), but tends to underbind as one moves away
from the valley of stability.

In this paper we analyze the ground-states and several excited states
of neutron drops for four Hamiltonians: AV8$^\prime$ without TNI,
AV8$^\prime$+UIX, AV8$^\prime$+IL7, and JISP16.  We examine possible
sub-shell closures and the spin-orbit splittings of odd systems near
closed HO shells.  We also compare results for the neutron matter
equations of state using AV8$^\prime$ with and without TNIs that could
be useful to calibrate bulk and gradient terms of Skyrme forces.  The
techniques we use are based on two quantum Monte Carlo (QMC)
techniques for the Argonne interactions and on the No-Core Full
Configuration (NCFC) for the nonlocal interaction JISP16.  We provide
quantified uncertainties where feasible.

The Green's function Monte Carlo (GFMC) provides accurate energies,
radii and and other properties of nuclei up to $A=12$ with the Argonne
interactions~\cite{Pieper:2008}; currently it can be used for systems
of up to  16 neutrons.  The Auxiliary Field Diffusion Monte Carlo
(AFDMC)~\cite{Schmidt:1999} has similar statistical accuracy as GFMC
in computing energies of systems of neutrons and
can be implemented for larger systems, up to more than 100
neutrons~\cite{Gandolfi:2009,Gandolfi:2010za}. A comparison between AFDMC
and GFMC results (obtained with the same Hamiltonian) suggests
that the systematic uncertainties are of the order of a few
percent. Improving the trial wave function used in AFDMC
to include pairing, for example, could further reduce these differences.

For JISP16 we expand the neutron drop wave functions in a HO basis.
For any finite truncation of the basis, this provides us with a strict
upper limit on the total energy of the system.  Exact results are
obtained by considering the limit of a complete (infinite-dimensional)
HO basis --- which we refer to as No-Core Full Configuration
(NCFC)~\cite{Maris:2008ax}.  We can obtain the total energies for
systems up to $A=14$ nucleons to within a percent by a simple
extrapolation to the complete basis from a series of successive finite
truncations.  The extrapolation of other observables such as radii and
densities is not as straightforward, but for small enough systems we
can simply consider a large enough basis space in order to obtain
converged results.  Note that in a single run, at a fixed truncation,
we not only obtain the ground state energy, but also the low-lying
spectrum of the system.

The plan of the paper is the following: in Sec.~\ref{sec:hamiltonians}
we describe various Hamiltonians we consider in this work.  Then, in
Sec.~\ref{sec:QMC} we briefly review the different Monte Carlo
many-body techniques used to solve for the neutron drops.
Sec.~\ref{sec:NCFC} presents an overview of the NCFC approach and
Sec.~\ref{sec:results} presents our main results for finite neutron
drops.  We present our results for neutron matter in
Sec.~\ref{sec:neutronmatter} and our conclusions in
Sec.~\ref{sec:conclusion}.

\section{Hamiltonians
\label{sec:hamiltonians}}

We adopt non-relativistic Hamiltonians with the following general
form:
\begin{equation}
H=-\sum_i{\hbar^2\over2m}\nabla_i^2
+\sum_i U_{\hbox{\scriptsize ext}}(r_i)+\sum_{i<j}v_{ij}+\sum_{i<j<k}V_{ijk} \,.
\end{equation} 
Systems consisting of only neutrons are not expected to be self-bound.
Therefore it is necessary to include an external well
$U_{\hbox{\scriptsize ext}}(r)$ in the Hamiltonian to have a confined
system.  We consider both harmonic oscillator (HO) wells and a
Woods-Saxon (WS) potential.

The HO wells have the form
\begin{equation}
U_{HO}(r)=\frac{1}{2}m\Omega^2r^2 \,.
\end{equation} 
This potential is useful due to its simplicity and the fact that the
ground state may be driven to arbitrary low density (i.e. with
arbitrarily weak external harmonic potential strength) or arbitrarily
large particle number.  The convergence of both the Monte Carlo and
the configuration interaction methods are improved due to the lack of
any low-lying states with long-range tails in the wave function.  This
feature enables applying our results for tests of EDFs over a range
from moderately low to rather high densities.  Most of our results are
for HO wells.  Woods-Saxon wells have been used in other calculations
of properties of neutron drops.  In particular neutron drops with a WS
well and the Argonne $v_{18}$ $N\!N$ and Illinois-2 potentials have
been shown to provide a good description of oxygen
isotopes~\cite{Pieper:2005}.  The WS form is
\begin{equation}
U_{WS}(r)=\frac{U_0}{1+e^{(r-R)/a}} \,,
\label{eq:ws-well}
\end{equation} 
where we have used $a=1.1$~fm, $U_0=-35.5$~MeV and $R=3$~fm, that is,
the same parameters as in Ref.~\cite{Chang:2004a}.

In addition to the total energy, $E = \langle H \rangle$, we also
calculate other observables, such as the external energy $\langle
U_{\hbox{\scriptsize ext}} \rangle$, the internal energy
$E_{\hbox{\scriptsize int}} = \langle H \rangle - \langle
U_{\hbox{\scriptsize ext}} \rangle$, and the rms radius $r$.  Note
that for a HO well, the external energy $\langle U_{\hbox{\scriptsize
    ext}} \rangle$ is proportional to $r^2$, and thus the quantities
$E$, $E_{\hbox{\scriptsize int}}$, and $r$ are not independent
observables in a HO well; however, in a WS well they are independent.

\subsection{Argonne $N\!N$ interaction and three-body forces
\label{sec:TNIs}}

One of the $N\!N$ potentials we adopt here is the Argonne
AV8$^\prime$~\cite{Pudliner:1997ck,Wiringa:2002}.  It is a simplified
version of the Argonne AV18~\cite{Wiringa:1994wb}, with the advantage
that it can be exactly included in both GFMC and AFDMC algorithms
without treating any part perturbatively. Other non-local operators
appearing in AV18 must be included as a perturbation of AV8$^\prime$
in QMC calculations~\cite{Pudliner:1997ck}.  These perturbative
corrections can be accurately computed within GFMC, but not in
AFDMC. Thus we consider AV8$^\prime$ to facilitate comparisons of the
two different QMC methods.  The difference between the binding
energies from AV8$^\prime$ and AV18 is very small in light nuclei, and
much smaller in pure neutron systems; about 0.06 MeV per
neutron~\cite{Pieper:2001}.

The Argonne AV8$^\prime$ is a sum of eight operators:
\begin{equation} 
v_{ij}=\sum_p v_p(r_{ij})O_{ij}^p
\end{equation} 
in which the $v_p(r_{ij})$ depend on the distance between the nucleons
$i$ and $j$, and $O_{ij}^p$ are operators.  Their form is
\begin{equation}
\label{eq:vop} 
O_{ij}^{p=1,8}=(1,\vec\sigma_i\cdot\vec\sigma_j,S_{ij},\vec
L_{ij}\cdot\vec S_{ij}) \times(1,\vec\tau_i\cdot\vec\tau_j) 
\end{equation}
with $S_{ij}$ the tensor operator, $\vec L_{ij}$ the relative angular
momentum and $\vec S_{ij}$ the total spin.  The $v_p(r)$ parts are fit to
reproduce the S and P partial waves as well as the $^3D_1$ wave and
its coupling to $^3S_1$ of the full AV18
potential~\cite{Wiringa:1994wb}.


In this paper we consider the AV8$^\prime$ alone and combined with two
different three-body forces: 
the Urbana-IX (UIX)~\cite{Pudliner:1995wk} 
and the Illinois-7 (IL7)~\cite{Pieper:2008}.  Just like for the $N\!N$
interaction, the TNIs are sums of several operators:
\begin{eqnarray}
V_{ijk} &=& A_{2\pi}^{PW} O^{2\pi,PW}_{ijk}
+ A_{2\pi}^{SW} O^{2\pi,SW}_{ijk}
+ A_{3\pi}^{\Delta R} O^{3\pi,\Delta R}_{ijk}
\nonumber \\ && {}
+A_R O^R_{ijk} + A_R^{T=3/2} O^{R,T=3/2}_{ijk}   \,.
\end{eqnarray}
Both TNIs include the Fujita-Miyazawa operator $O^{2\pi,PW}$ and a
phenomenological part $O^R$, while only IL7 has the
$O^{2\pi,SW}_{ijk}$ ,$O^{3\pi,\Delta R}_{ijk}$, and
$O^{R,T=3/2}_{ijk}$ terms.  In the Fujita-Miyazawa
term~\cite{Fujita:1957} two pions are exchanged between the three
nucleons with the creation of an intermediate excited state.  The
phenomenological part $O^R$ has no spin or isospin dependence.  The
additional IL7 terms involve exchanges of two or three pions and a
pure $T=3/2$ repulsion.  A full description of the operators is given
in Refs.~\cite{Pieper:2001,Pieper:2008}.  

The $A_{2\pi}^{PW}$ and $A_R$ parameters of UIX are determined by
reproducing the binding energy of $^3$H and nuclear
matter~\cite{Pudliner:1995wk}.  The UIX model has been used to
investigate properties of neutron matter (see for example
Refs.~\cite{Akmal:1998,Gandolfi:2009,Gandolfi:2012} and references
therein).  The resulting equation of state will support neutron stars
larger than two solar masses.  

The Illinois forces are more sophisticated than UIX.  The $A$
coefficients are determined by fits to binding energies of light
nuclei~\cite{Pieper:2001,Pieper:2008}.  The Illinois forces give a
good description of properties of nuclei up to $A=12$, including both
ground states and excited states, however three-pion operators are
very attractive in pure neutron systems and they overbind neutron
matter at large densities~\cite{Sarsa:2003}.  In this work, we
consider IL7 as described in~\cite{Pieper:2008}.

\subsection{JISP16}

The JISP16 $N\!N$ interaction is determined by inverse scattering
techniques from the $np$ phase shifts and is, therefore, charge
symmetric.  JISP16 is available in a relative HO
basis~\cite{Shirokov07} and can be written as a sum over partial waves
\begin{equation}
 {\hat V} = \sum_{S, {\cal J}, T} {\cal P}_{S, {\cal J}, T} 
 \sum_{n, \ell, n^\prime, \ell^\prime}
 \;| n l\rangle 
 \
 A^{S, {\cal J}, T}_{n\ell,n^\prime \ell^\prime}
 \
 \langle n^\prime \ell^\prime | 
\end{equation}
where $\hbar \Omega = 40$ MeV and ${\vec {\cal J}} = {\vec \ell} +
{\vec s}$.  The HO relative coordinate wave function is written
$\langle r | n \ell \rangle = {\cal R}_{n \ell} (r) $.  A small number
of coefficients $\{ A^{S, {\cal J}, T}_{n\ell,n^\prime \ell^\prime}\}$
are sufficient to describe the phase shifts in each partial wave.
Note that the JISP16 interaction is non-local and its off-shell
properties have been tuned by phase-shift equivalent transformations
to produce good properties of light nuclei.  For example, JISP16 is
tuned in the $^3S_1-^3D_1$ channel to give a high precision
description of the deuteron's properties.  Other channels are tuned to
provide good descriptions of $^3$H binding, the low-lying spectra of
$^6$Li and the binding energy of $^{16}$O~\cite{Shirokov07}.  After
its initial introduction, it was realized that the $^{16}$O energy was
not fully converged and JISP16 overbinds $^{16}$O by about 15 to 20
MeV~\cite{Maris:2008ax}.  With these off-shell tunings to nuclei with
$A \geq 3$ one may view JISP16 as simulating, to some approximation,
what would appear as $N\!N\!N$ interaction contributions (as well as
higher-body interactions) in alternative formulations of the nuclear
Hamiltonian.

\section{Quantum Monte Carlo methods
\label{sec:QMC}}

Both of our QMC methods use diffusion Monte Carlo to project the
lowest-energy eigenstate out of a trial wave function $\Psi_T$ by a
propagation in imaginary time $\tau$:
\begin{equation}
\Psi(\tau)=e^{-(H-E_T)\tau}\Psi_T \,,
\label{eq:qmc-prop}
\end{equation}
where $E_T$ is a normalization factor.  In the $\tau\rightarrow\infty$
limit the only component of $\Psi_T$ that survives is the
lowest-energy one not orthogonal to $\Psi_T$:
\begin{equation}
\Psi_0=\lim_{\tau\rightarrow\infty}\Psi(\tau) \,.
\end{equation}
The evolution in imaginary time is performed by using Monte Carlo
integration to evaluate
\begin{equation}
\Psi(R,\tau)=\int dR^\prime G(R,R^\prime,\tau) \Psi_T(R^\prime) \,,
\end{equation}
where $G(R,R^\prime,\tau)$ is an approximation of the many-body
Green's function of the Hamiltonian, and $R$ and $R^\prime$ are the
positions of all $A$ nucleons: $ R\equiv (\vec r_1,\dots,\vec r_N)$. 
The exact form of $G(R,R^\prime,\tau)$ is unknown, but it can be
accurately approximated as a product of many
$G(R,R^\prime,\Delta\tau)$ for a small time step, $\Delta\tau$.  
The main difference between GFMC and AFDMC is in their representations
of $\Psi$ and the structure of the initial $\Psi_T$.

\subsection{GFMC method and trial wavefunction}

GFMC uses a complete spin-isospin representation of the many-body wave
function; $\Psi(R,\tau)$ is written as a vector with $2^A N(T)$
complex components.  Here the $2^A$ allows for all possible nucleon
spin up or down combinations and $N(T)$ is the number of
proton-neutron combinations with the desired isospin.  In the case of
neutron drops $N(T)=1$.  The exponential growth of the vector size
with the number of neutrons currently limits GFMC calulations of
neutron systems to $N \leq 16$.

Calculations of nuclei with realistic interactions face a sign problem
eventually as the trial wave function is propagated to large $\tau$.
To deal with this problem, we use the constrained path algorithm to
obtain configurations with the largest possible overlap with the
ground state~\cite{Wiringa:2000}.  This method is similar to the fixed
node approximation in that it is exact in the limit of an exact
constraint, and it is stable to large imaginary time; however it does
not provide an upper bound.  We then extend the propagation without
constraint for as long as possible to obtain the energy and
ground-state properties. The constraint and the
convergence properties of GFMC are discussed in~\cite{Wiringa:2000},
Figure 3 in that paper shows some convergence results for neutron
drops.

The trial wave functions used in our GFMC calculations for neutron
drops in HO wells are somewhat simplified from the ones described in
Ref.~\cite{Pudliner:1997ck} for nuclei:
\begin{eqnarray}
     |\Psi_T\rangle &=& \left[ {\cal S}\prod_{i<j}(1+U_{ij}) \right] 
                      |\Psi_J\rangle \ ,
\label{eq:psit}
\\
     |\Psi_J\rangle &=& \left[ \prod_{i<j}f_c(r_{ij}) \right] 
                     |\Phi_N(JMTT_{3})\rangle \ .
\label{eq:jastrow}
\end{eqnarray}
The non-central $U_{ij}$ and associated central $f_c$ are optimal
correlations for neutron matter of the form described in
Ref.~\cite{Pieper:1992}.  For drops with $N \leq 8$, the $\Phi_N$ is
expanded in an $LS$ basis of $s$- and $p$-shell oscillator functions
as described in Ref.~\cite{Pudliner:1997ck}.  For $N \geq 8$ we use a
``BCS'' ansatz $\Phi_{\hbox{\scriptsize BCS}}$ of the form introduced in
Refs.~\cite{Carlson:2003,Chang:2004}.  For $N=8$ the two forms give
very similar variational energies.

The BCS pairing is important, particularly for low-density systems and
when trying to calculate even-odd staggering of the energies.  In this
paper we consider $\Phi_{\hbox{\scriptsize BCS}}$ for only $J=0$
states or, in the case of odd $N$, for states in which the total $J$
is carried by a single neutron.  Such $\Phi_{\hbox{\scriptsize BCS}}$
are written using correlated pairs of neutrons with total spin $S=0$
and a $L=0$ spatial wave function $\phi_{ij}$ expanded in 0$s$, 0$p$,
1$s$, and 0$d$ single-neutron wave functions:
\begin{eqnarray}
  \phi_{ij} &=& \beta_{0s} \phi_{0s}(r_i)\phi_{0s}(r_j) +  \beta_{0p} \phi_{0p}(r_i)\phi_{0p}(r_j)P_1(\hat{r}_{ij}) \nonumber \\ 
     && {} +  \beta_{1s} \phi_{1s}(r_i)\phi_{1s}(r_j)  \nonumber \\ 
     && {} +  \beta_{0d} \phi_{0d}(r_i)\phi_{0d}(r_j)P_2(\hat{r}_{ij})  \ .
\label{eq:bcs-phi}
\end{eqnarray}
The $\beta_{nl}$ are variational parameters; only their ratios are relavant.
For even $N$, 
\begin{eqnarray}
\Phi_{\hbox{\scriptsize BCS}}(J=0) = \sum_{P} | \phi_{ij} | ~~| s_1s_2\cdots s_N \rangle  \ ,
\label{eq:bcs-even-n}
\end{eqnarray}
where the sum is over all partitions of $N$ neutrons into $N/2$ spin-up neutrons and 
$N/2$ spin-down neutrons, $i$ is from the set of spin-up neutrons and $j$ is from
the set of spin-down neutrons, $| \phi_{ij} |$ is a determinant, 
and $| s_1s_2\cdots s_N \rangle$ is the spin-vector component with the given spins.
For odd $N$ and $s_{1/2}$ or $d_{5/2}$ states we have
\begin{equation}
\Phi_{\hbox{\scriptsize BCS}}(M\!=\!J) =  {\cal A} \left[ \phi_{L,S,J,M} ({\bf r}_{N},\sigma_N) \ \Phi_{\hbox{\scriptsize BCS,N-1}}(J\!=0\!) \right]   \ ,
\label{eq:bcs-odd-n}
\end{equation}
which is also a sum over partitions of determinants.

\begin{figure}
\center\includegraphics[height=0.99\columnwidth,angle=270]{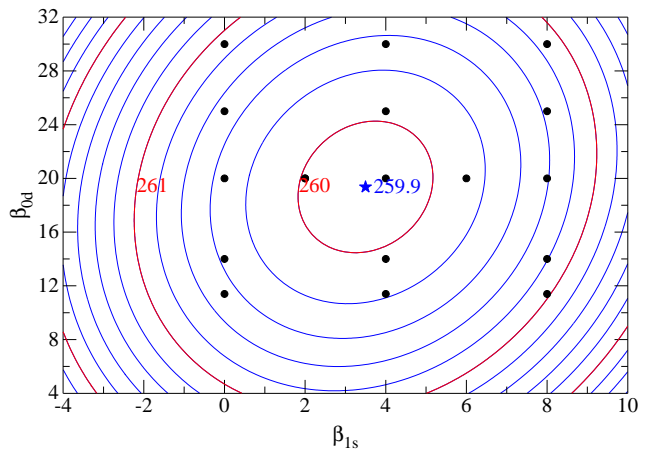}
\caption{(color online) Contours of VMC energies for 12 neutrons in a
  10-MeV HO well with AV8$^\prime$+UIX versus the 0$d$ and 1$s$
  coefficients in the BCS wave function.  The dots show the cases that
  were computed using GFMC.
  \label{fig:bcs_contours}}
\end{figure}
Figure~\ref{fig:bcs_contours} shows the variational Monte Carlo (VMC)
energies computed for systems of 12 neutrons with different BCS
parameters for the 0$d$ and 1$s$ pairs; the $\beta_{0s}$ and $\beta_{0p}$ 
are fixed at 100 each so the $^8$n core is almost full.  
The optimized choice is used for the GFMC calculations.
The total VMC or GFMC energy is not very sensitive to the choice of
parameters (the contour interval is only 0.08\%) - the effects on pairing can be larger.  
A detailed description of the algorithm as well as the importance sampling
technique (constrained propagation) used to reduce the variance can be found in
Refs.~\cite{Pudliner:1997ck,Wiringa:2000}.

The VMC energies computed with $\Psi_{T}$ or even just $\Psi_J$ are much
closer to the final GFMC energies than is the case for real nuclei.
The values using $\Psi_J$ are typically less than 10\% above the
GFMC values and often display better convergence as the number of unconstrained
steps (see Ref.~\cite{Wiringa:2000}) is increased.

\subsection{GFMC numerical convergence and error estimate}

QMC calculations have an easily quantified statistical error arising
from the Monte Carlo method.  It is not so straightforward to
determine the magnitude of the systematic errors arising from
approximations made in the constrained-path implementation.  These
have been extensively discussed for GFMC in
Refs.~\cite{Pudliner:1997ck,Wiringa:2000}.

Quantities other than the energy are usually evaluated by combining
mixed and variational estimates:
\begin{equation}
\langle \Psi_0 | {\cal O} | \Psi_0 \rangle =
2 \langle \Psi_0 | {\cal O} | \Psi_T \rangle -
\langle \Psi_T | {\cal O} | \Psi_T \rangle.
\end{equation}
No extrapolation for the energy is required since the ground state is
an eigenstate of the Hamiltonian and the propagation commutes with the
Hamiltonian.  This linear extrapolation is usually very accurate if
the calculation begins with a good trial wave function. It is also possible to
estimate other observables by forward walking techniques or by adding
a small perturbation $\epsilon {\cal O}$ to the Hamiltonian and
computing the difference between the energy of the original and
perturbed Hamiltonian. We have not pursued these methods for the
present paper. Hence the internal energy results as well as other
quantities reported below are extrapolated and potentially not as
accurate as the full energy.

\subsection{AFDMC method and trial wave function
\label{sec:afdmc_wf}}

The presence of spin operators in the Hamiltonian requires a summation
of all possible good spin states in the wave function.  In AFDMC the
spin states are sampled using Monte Carlo
techniques~\cite{Schmidt:1999}.  This sampling is performed by
reducing the quadratic dependence of spin operators in the exponential
of Eq.~(\ref{eq:qmc-prop}) to a linear form by means of a
Hubbard-Stratonovich 
transformation.  The effect of an exponential of a linear combination
of spin operators consists of a rotation of the spinor for each
neutron during the propagation, and this permits the use of a much
simpler basis in the trial wave function.  The result is that the wave
function is less accurate, but it can be rapidly evaluated.  Since
both positions and spins can be sampled, the AFDMC method can be used
to solve for the ground state of much larger systems than GFMC -- more
than one-hundred neutrons currently may be solved with AFDMC.

More detailed explanations of the AFDMC method and how to include the
full $N\!N$ interactions and TNIs in the propagator can be found in
Refs.~\cite{Sarsa:2003,Pederiva:2004,Gandolfi:2009},
where the constrained-path approximation used to control the fermion
sign problem is also discussed.

The AFDMC method projects out the lowest energy state with the same
symmetry as the trial wave function from which the projection is
started.  The trial wave function used in the AFDMC algorithm has the
form:
\begin{eqnarray} 
|\Psi_T (R,S)\rangle=\left[\prod_{i<j}f(r_{ij})\right] 
|\Phi_{JM}(R,S)\rangle \,,
\label{eq:jasafdmc}
\end{eqnarray}
where $S\equiv (s_1,\dots ,s_N)$.
The spin assignments $s_i$ consist in giving the spinor components, namely
\begin{equation}
s_i \equiv \left(\begin{array}{c} 
u_i \\ d_i
\end{array}\right)=u_i |\uparrow\rangle + d_i |\downarrow\rangle \ , 
\end{equation}
where $u_i$ and $d_i$ are complex numbers.  The Jastrow function
$f(r)$ is the solution of a Schr\"odinger-like equation for $f(r<d)$,
\begin{equation}
-\frac{\hbar^2}{m}\nabla^2f(r)+\alpha v_c(r)f(r)=\lambda f(r) \,,
\end{equation}
where $v_c(r)$ is the spin-independent part of the nucleon-nucleon
interaction, and the healing distance $d$ and $\alpha$ are variational
parameters.  For distances $r\ge d$ we impose $f(r)=1$.  The Jastrow
part of the function in our case has the only role of reducing the
overlap of nucleons, therefore reducing the energy variance.

The antisymmetric part of the wave function is
\begin{eqnarray} 
\Phi_{JM}(R,S) = \Big[\sum D\{\phi_\alpha(\vec r_i,s_i)\}\Big]_{J,M} \,,
\end{eqnarray}
where $\alpha=\{n,l,j,m_j\}$ is the set of quantum numbers of
single-particle orbitals, and the summation of determinants gives a
trial wave function that is an eigenstate of $J^2$ and $M$.  The
single-particle basis is given by
\begin{eqnarray}
\phi_\alpha(\vec r_i,s_i)=\Phi_{nlj}(r_i)\left[Y_{l,m_l}(\hat r_i)\xi_{s,m_s}(s_i)\right]_{j,m_j} \,.
\end{eqnarray}
The radial components $\Phi_{nlj}$ are obtained by solving the
Hartree--Fock problem with the Skyrme force SKM~\cite{Pethick:1995},
$Y_{l,m_l}$ are spherical harmonics, and $\xi_{s,m_s}$ are spinors in
the usual up-down basis.  For each $(J,M)$ set of quantum numbers
there are several combinations of single-particle orbitals.  We
typically perform several simulations to identify the ground-state and
order of excited states.  It is also possible to include a BCS pairing
term in the trial wave function in AFDMC, as has been done for GFMC.
This would allow more accurate treatment of pairing in very
low-density systems, and is currently under development.

For neutron matter calculations we change the antisymmetric part of
the wave function to be the ground state of the Fermi gas, built from
a set of plane waves.  The infinite uniform system is simulated by a
cubic periodic box of volume $L^3$ according to the density of the
system. The momentum vectors in this box are
\begin{equation}
\bold k_\alpha=\frac{2\pi}{L}(n_{\alpha x},n_{\alpha y},n_{\alpha z}) \,,
\end{equation}
where $\alpha$ labels the quantum state and $n_x$, $n_y$ and $n_z$ are
integer numbers describing the state.  The single-particle orbitals
are given by
\begin{equation}
\phi_\alpha(\vec r,s)=e^{i\vec k_\alpha\cdot\vec r}
\xi_{s,m_s,\alpha}(s) \,.
\end{equation}
Again, it is possible to generalize the neutron matter calculations to
include BCS pairing in the trial state~\cite{Gandolfi:2008b,Gandolfi:2009b}.

\subsection{AFDMC numerical convergence and error estimate
\label{sec:afdmc_conv}}

AFDMC is very similar in concept to GFMC and convergence and error
estimates are also similar.  The statistical errors are easily
evaluated and controllable.  AFDMC also uses a constrained-path method
to circumvent the fermion sign problem, and hence the results depend
to some degree on the choice of trial function.

Although it is possible to do some unconstrained propagation with
AFDMC, it is more limited than GFMC because the sampling of the spins
introduces a fermion sign problem earlier.  Uncertainties can also be
addressed by choosing several different initial trial states.  We find
the constrained path results to be reasonably accurate when compared
with GFMC for small systems.

\begin{table}
\renewcommand{\arraystretch}{1.2}
\begin{ruledtabular}
\begin{tabular}{rr|dd|dd}
    &  & \multicolumn{1}{c}{GFMC} 
       & \multicolumn{1}{c|}{AFDMC} & \multicolumn{2}{c}{Difference}\\
    &  & \multicolumn{1}{c}{(MeV)}
       & \multicolumn{1}{c|}{(MeV)} 
       & \multicolumn{1}{c}{(MeV)}   
       & \multicolumn{1}{c}{ \% }  \\
\hline
 $N$ & $J^\pi$ & \multicolumn{4}{c}{5 MeV HO well}\\
\hline
 8 &    0$^+$         &  67.00(1) &  67.1(1) & $0.1(1)$ & $0.1$ \\
 9 &  $\frac{1}{2}^+$ &  80.90(4) &  81.2(1) & $0.3(1)$ & $0.4$ \\
 9 &  $\frac{5}{2}^+$ &  81.20(3) &  81.9(1) & $0.7(1)$ & $0.9$ \\
10 &    0$^+$         &  92.1(1)  &  94.6(1) & $2.5(2)$ & $2.7$ \\
11 &  $\frac{5}{2}^+$ & 106.3(1)  & 108.0(1) & $1.7(2)$ & $1.6$ \\
11 &  $\frac{1}{2}^+$ & 105.9(1)  & 108.4(1) & $2.5(2)$ & $2.3$ \\
12 &    0$^+$         & 118.1(1)  & 121.1(1) & $3.0(2)$ & $2.5$ \\
13 &  $\frac{5}{2}^+$ & 131.5(1)  & 135.7(2) & $4.2(3)$ & $3.1$ \\
13 &  $\frac{1}{2}^+$ & 130.8(1)  & 134.1(2) & $3.3(3)$ & $2.5$ \\
14 &    0$^+$         & 142.2(2)  & 146.7(1) & $4.5(3)$ & $3.1$ \\
\hline
 $N$ & $J^\pi$ & \multicolumn{4}{c}{10 MeV HO well}\\
\hline
 3 &  $\frac{3}{2}^-$ &  45.5(0)  &  45.0(1) & $-0.5(1)$ & $-1.1$ \\
 3 &  $\frac{1}{2}^-$ &  46.70(1) &  46.7(1) & $ 0.0(1)$ & $ 0.0$ \\
 4 &    0$^+$         &  62.00(1) &  62.9(1) & $ 0.9(1)$ & $ 1.4$ \\
 5 &  $\frac{3}{2}^-$ &  83.00(1) &  82.9(1) & $-0.1(1)$ & $-0.1$ \\
 5 &  $\frac{1}{2}^-$ &  84.00(2) &  83.7(1) & $-0.3(1)$ & $-0.3$ \\
 6 &    0$^+$         &  98.90(2) &  98.4(1) & $-0.5(1)$ & $-0.5$ \\
 7 &  $\frac{1}{2}^-$ & 118.9(0)  & 118.0(1) & $-0.9(1)$ & $-0.7$ \\
 7 &  $\frac{3}{2}^-$ & 121.1(0)  & 120.6(1) & $-0.5(1)$ & $-0.4$ \\
 8 &    0$^+$         & 135.8(0)  & 134.7(1) & $-1.1(1)$ & $-0.8$ \\
 9 &  $\frac{1}{2}^+$ & 163.7(1)  & 163.5(1) & $-0.2(2)$ & $-0.1$ \\
 9 &  $\frac{5}{2}^+$ & 163.2(1)  & 162.5(1) & $-0.7(2)$ & $-0.4$ \\
10 &    0$^+$         & 188.1(6)  & 188.5(1) & $ 0.4(7)$ & $ 0.2$ \\
11 &  $\frac{5}{2}^+$ & 217.0(3)  & 216.7(1) & $-0.3(4)$ & $-0.1$ \\
11 &  $\frac{1}{2}^+$ & 216.1(3)  & 216.6(2) & $ 0.5(5)$ & $ 0.2$ \\
12 &    0$^+$         & 242.0(6)  & 240.8(1) & $-1.2(7)$ & $-0.5$ \\
13 &  $\frac{5}{2}^+$ & 267.6(6)  & 266.3(2) & $-1.3(8)$ & $-0.5$ \\
13 &  $\frac{1}{2}^+$ & 268.0(5)  & 267.2(2) & $-0.8(7)$ & $-0.3$ \\
14 &    0$^+$         & 291.9(2)  & 291.2(2) & $-0.7(4)$ & $-0.2$ \\
\end{tabular}
\end{ruledtabular}
\caption{Comparison of GFMC and AFDMC total energies for neutron drops
  in 5 MeV and 10 MeV wells with AV8$^\prime$+UIX.  Statistical errors
  due to the Monte Carlo sampling are given in brackets.
\label{tab:gfmc-afdmc}}
\end{table}
In Table~\ref{tab:gfmc-afdmc} we compare the GFMC and AFDMC total
energy results for neutron drops in 5 and 10 MeV HO wells.  Overall
the agreement is of the order of a few percent or better.  Statistical
errors due to sampling can be made arbitrarily small, these are about
0.2\% or less for the AFDMC and GFMC calculations.

For the lower densities generated by the 5 MeV well the GFMC energies
are significantly lower than the AFDMC energies, by up to 3\%.  A
plausible explanation for these discrepancies is the fact that we have
not incorporated BCS pairing in the AFDMC calculation.  Indeed, the
differences are nearly zero for the closed shell at $N=8$ (where
pairing does not play a role) and grow as we go towards the middle of
the shell, $N=14$.  We are therefore pursuing the inclusion of BCS
pairing in the AFDMC calculations of neutron drops in order to improve
the results at low densities.  On the other hand, for the 10 MeV well
the GFMC and AFDMC results are all within 1\% of each other, with the
AFDMC typically lower than the GFMC energies.

Systematic errors in calculations of neutron matter are similar in
spirit.  The trial wave function can affect results for the energy at
low densities where pairing is important.  At larger densities,
though, pairing provides a very small fraction of the total energy of
the system and calculations are much less sensitive to the choice of
the trial state.

In addition to the Monte Carlo errors and the dependence on the choice
of the trial state, we have to consider finite-size effects for
calculations of neutron matter.  We enforce periodic boundary
conditions and fix the number of neutrons to be a closed shell in the
periodic free-particle basis.  In order to reduce finite-size effects
we performed simulations with 66 neutrons.  Free fermions for $N=66$
provide a kinetic energy very close to the infinite limit.  Any
possible finite-size effect due to the truncation of the potential
energy is properly taken into account by considering several replicas
of the simulation box as described in
Ref.~\cite{Sarsa:2003,Gandolfi:2009}.  Typically these uncertainties
are very small for bulk properties like the ground-state energy.

\section{No Core Full Configuration method
\label{sec:NCFC}}

The NCFC method is based on a series of no-core configuration
interaction calculations with increasing basis dimensions.  In this
approach the wave function for the $N$ neutrons is expanded in an
$N$-body basis of Slater determinants of single-particle states, and
the many-body Schr\"odinger equation becomes a large sparse matrix
eigenvalue problem.  We obtain the lowest eigenstates of this matrix
iteratively.  In a complete basis, this method would give exact
results for a given input interaction $V$.  However, practical
calculations can only be done in a finite-dimensional truncation of a
complete basis.  We perform a series of calculations until we reach
numerical convergence in a sufficiently large basis space, or we
employ a simple extrapolation~\cite{Maris:2008ax} to the complete
basis.

\subsection{Description of basis space}

Our choice for the basis is the harmonic oscillator (HO) basis so
there are two basis space parameters, the HO energy $\hbar\omega$ and
the many-body basis space cutoff $N_{\hbox{\scriptsize max}}$.  The
cutoff parameter $N_{\hbox{\scriptsize max}}$ is defined as the
maximum number of total oscillator quanta allowed in the many-body
basis space above the minimum for that number of neutrons.  Numerical
convergence is defined as independence of both basis space parameters
$N_{\hbox{\scriptsize max}}$ and $\hbar\omega$, within evaluated
uncertainties.  Note that the basis space parameter $\hbar\omega$ is
not necessarily the same as that of the HO well $\hbar\Omega$ that
confines the neutrons.

We employ a many-body basis in the so-called $M$-scheme: the many-body
basis states are Slater determinants in a HO basis, limited by the
imposed symmetries --- parity and total angular momentum projection
($M$), as well as by $N_{\hbox{\scriptsize max}}$.  Each
single-particle HO state has its orbital and spin angular momenta
coupled to good total angular momentum, $j$, and magnetic projection,
$m$.  Here we only consider natural-parity states, and utilize $M=0$
for an even number of neutrons, and $M=\frac{1}{2}$ for an odd number
of neutrons.  In this scheme a single calculation gives the entire
spectrum for that parity and $N_{\hbox{\scriptsize max}}$.

The NCFC approach satisfies the variational principle and guarantees
uniform convergence from above to the exact eigenvalue with
increasing $N_{\hbox{\scriptsize max}}$.  That is, the results for the
energy of the lowest state of each spin and parity, at any
$N_{\hbox{\scriptsize max}}$ truncation, are strict upper bounds on
the exact converged answers and the convergence is monotonic with
increasing $N_{\hbox{\scriptsize max}}$.

The challenge for this approach is that the matrix dimension grows
nearly exponentially with increasing $N_{\hbox{\scriptsize max}}$.
The calculations presented here have been performed with the code
MFDn~\cite{DBLP:conf/sc/SternbergNYMVSL08,DBLP:journals/procedia/MarisSVNY10,DBLP:conf/europar/AktulgaYNMV12}
which has been demonstrated to scale to over 200,000 cores.  For small
neutron drops we are able to achieve converged results to within a
fraction of a percent by using a sufficiently large basis space, at
least for neutrons in a HO well of 10 MeV and above.  In order to
achieve converged NCFC results directly (i.e. without extrapolation)
for more than 10 neutrons using JISP16, we would need to obtain
eigenstates of matrices that are beyond the reach of present
technologies.  However, for up to 22 neutrons, we can utilize a
sequence of results obtained with $N_{\hbox{\scriptsize max}}$ values
that are currently accessible, in order to extrapolate to the infinite
or complete basis space limit.

\subsection{NCFC numerical convergence and error estimate}

We carefully investigate the dependence of the results on the basis
space parameters, $N_{\hbox{\scriptsize max}}$ and $\hbar\omega$.  Our
goal is to achieve independence of both of these parameters as that is
a signal for convergence --- the result that would be obtained from
solving the same problem in a complete basis.  For the total energy,
$\langle H \rangle$, the guarantee of monotonic convergence from above
to the exact total energy facilitates our choice of extrapolating
function.

We use an extrapolation method that was found to be reliable in light
nuclei: a constant plus an exponential in 
$N_{\hbox{\scriptsize max}}$~\cite{Forssen:2008qp,Maris:2008ax}.  
That is, for each set of three successive $N_{\hbox{\scriptsize max}}$ 
values at fixed $\hbar\Omega$, we fit the ground state energy with
three adjustable parameters using the relation
\begin{equation}
E_{gs}(N_{max}) = a \exp(-c\,N_{\hbox{\scriptsize max}}) + E_{gs}({\infty}) \,.
\label{extreq}
\end{equation}
Under the assumption that the convergence is indeed exponential, 
such an extrapolation should get more accurate as
$N_{\hbox{\scriptsize max}}$ increases; we use the difference between
the extrapolated results from two consecutive sets of three
$N_{\hbox{\scriptsize max}}$ values as an estimate of the numerical
uncertainty associated with the extrapolation.

\begin{figure}[t]
\center\includegraphics[width=0.9\columnwidth]{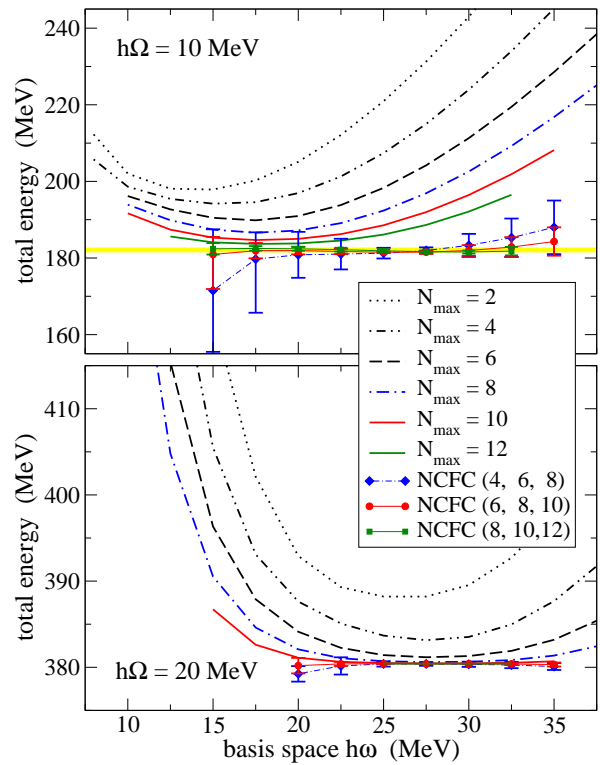}
\caption{(color online) 
  Ground state energy for 10 neutrons in a HO well of 10 MeV (top)
  and 20 MeV (bottom) for a series of finite basis space calculations
  with JISP16.  Note that our results in the 20 MeV well converge much
  more rapidly than in the 10 MeV well.  The symbols represent
  extrapolated results as discussed in the text and the error bars
  signify our uncertainty estimate at that basis space $\hbar\omega$;
  the yellow band represents our final result including numerical
  error estimates.
\label{fig:NCFCconvergence}}
\end{figure}
\begin{figure*}[t]
\center{
\includegraphics[width=0.9\columnwidth]{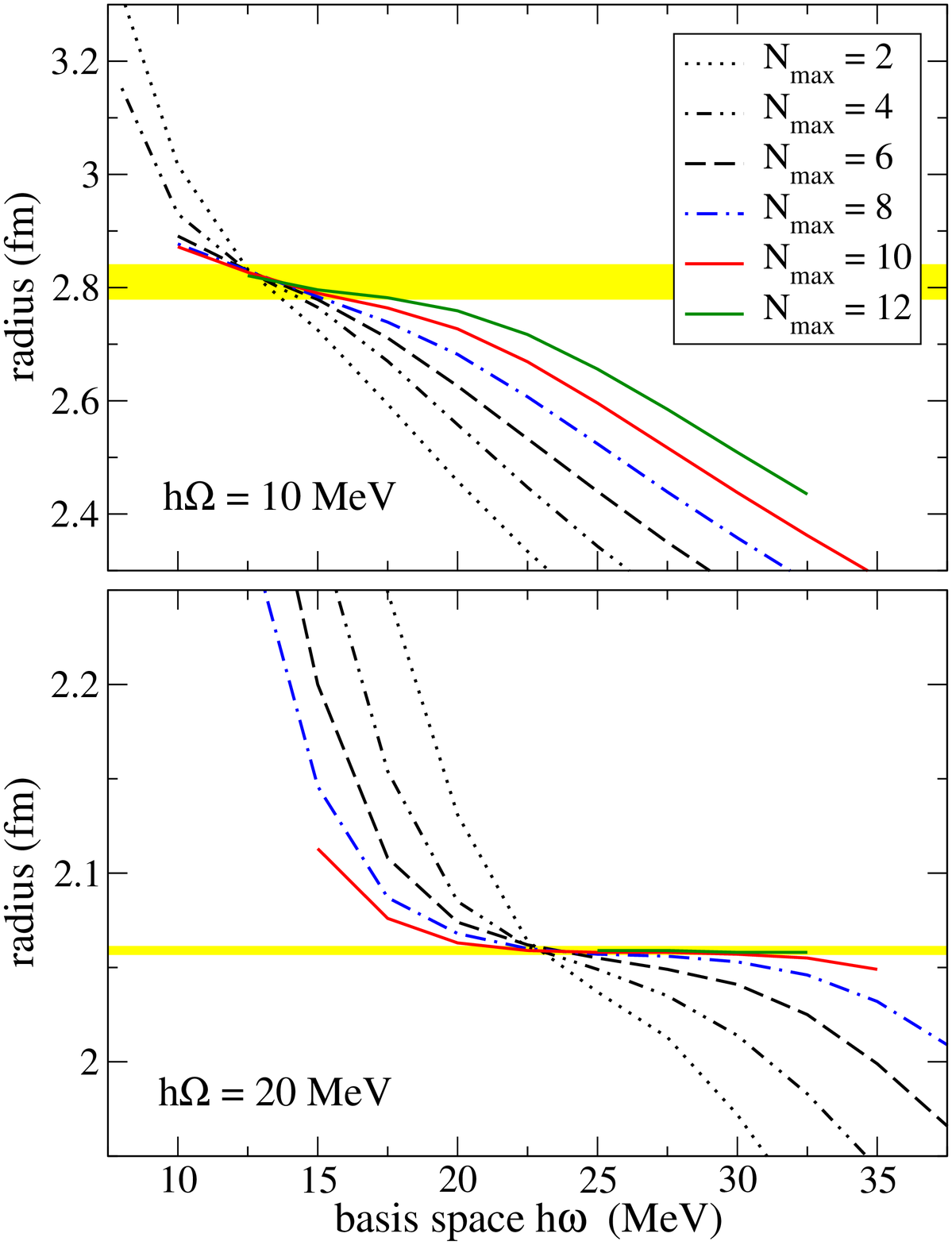}\qquad
\includegraphics[width=0.9\columnwidth]{res_JISP16_N10_internal}}
\caption{(color online) 
  The rms radius (left) and internal energy (right) for 10 neutrons in
  a HO well of 10 MeV (top) and 20 MeV (bottom) for a series of finite
  basis space calculations with JISP16.  Note that our results in the
  20 MeV well converge much more rapidly than in the 10 MeV well.  The
  yellow band represents our best estimate (including an error
  estimate) for the infinite basis space result.
\label{fig:NCFCconv_other}}
\end{figure*}
For a reasonable range of basis space parameters $\hbar\omega$, this
assumption appears to be valid, as is evident from
Fig.~\ref{fig:NCFCconvergence}.  In this figure we show the ground
state energies for 10 neutrons in a 10 MeV and 20 MeV HO well for a
series of finite bases, as well as the extrapolations with their
uncertainties indicated by error bars.  The error bars on the
extrapolations using calculations up to $N_{\hbox{\scriptsize max}} =
10$ are all smaller than the corresponding error bars from
calculations up to $N_{\hbox{\scriptsize max}} = 8$; and the error
bars on the extrapolations using calculations up to
$N_{\hbox{\scriptsize max}} = 12$ are all smaller than the
corresponding error bars from $N_{\hbox{\scriptsize max}} = 10$.
Furthermore, we see that the dependence on the basis space
$\hbar\omega$ decreases with increasing $N_{\hbox{\scriptsize max}}$,
and that the extrapolated results for a given $N_{\hbox{\scriptsize
    max}}$ agree within each other's error bars.  Our total error
estimate is based on a 5 MeV region in $\hbar\omega$ which has the
smallest error bars and minimal $\hbar\omega$ dependence.

In order to perform this extrapolation to the infinite basis space, 
we need finite basis space calculations up to 
$N_{\hbox{\scriptsize max}}=8$ or higher.  Above 22 neutrons, our
calculations are limited to $N_{\hbox{\scriptsize max}}=4$, so we only
have variational upper bounds for the total energy of these larger
neutron drops.  For the 10 MeV HO well, our results are not yet
converged at this basis space, but for the 20 MeV HO well, these
upper bounds are likely to be within a few percent of the converged
results.

The internal energies and the rms radii do not converge monotonically
with $N_{\hbox{\scriptsize max}}$, in contrast to the total energy.
Currently, we do not have a reliable method to perform the
extrapolation to the infinite basis space for these observables.  The
rms radius seems to converge from above for small basis space
parameters $\hbar\omega$, but from below for large basis space
parameters, see the left panels of Fig.~\ref{fig:NCFCconv_other}.
Hence there is a 'sweet spot' in the basis space parameter
$\hbar\omega$ for which the radius and equivalently, the external
energy $\langle U_{\hbox{\scriptsize ext}}\rangle$, is approximately
independent of $N_{\hbox{\scriptsize max}}$.  We use our results in
this region as an estimate of the infinite basis space result, with
error bars based on the residual $N_{\hbox{\scriptsize max}}$ and
$\hbar\omega$ dependence in a window around the 'sweet spot', as
indicated by the yellow band in the left panels of
Fig.~\ref{fig:NCFCconv_other}.

The internal energy appears to converge from above, at least for
$\hbar\omega$ values in the region that is optimal for the total and
external energies.  However, we have not been able to find a robust
convergence pattern; e.g. an exponential extrapolation does not work
very well, as can be seen in the top right panel of
Fig.~\ref{fig:NCFCconv_other}.  The most reliable estimate for the
internal energy in the infinite basis space appears to be the
difference between the (extrapolated) total energy and the external
energy based on the convergence at the 'sweet spot' explained above,
$E_{\hbox{\scriptsize int}} = \langle H \rangle - \langle
U_{\hbox{\scriptsize ext}} \rangle$.  This is depicted by the yellow
band in the right panels of Fig.~\ref{fig:NCFCconv_other}.

In order to make a meaningful estimate of the converged (NCFC)
results for the rms radius and the external and internal energies, 
we need to perform a set of calculations for a range of basis space
parameters $\hbar\omega$ at least up to 
$N_{\hbox{\scriptsize max}}=10$.  We therefore give these results only
up to 14 neutrons.

\section{Results for neutron drops
\label{sec:results}}

\subsection{Total energy}

\begin{table*}
\renewcommand{\arraystretch}{1.2}
\begin{ruledtabular}
\begin{tabular}{rr|dd|dd|dd|d}
     &        & \multicolumn{2}{c|}{ 5 MeV HO well} 
              & \multicolumn{2}{c|}{10 MeV HO well} 
              & \multicolumn{2}{c|}{20 MeV HO well} 
              & \multicolumn{1}{c}{WS well}  \\ 
\hline
 $N$ & $J^\pi$ & \multicolumn{1}{c}{AV8$^\prime$+UIX} & \multicolumn{1}{c|}{JISP16}
              & \multicolumn{1}{c}{AV8$^\prime$+UIX} & \multicolumn{1}{c|}{JISP16}
              & \multicolumn{1}{c}{AV8$^\prime$+UIX} & \multicolumn{1}{c|}{JISP16} 
              & \multicolumn{1}{c}{AV8$^\prime$+UIX} \\
\hline
 3 & $\frac{1}{2}^-$ &  22.89     & 22.73(1) & 46.69(1) & 46.512    &  97.1(1)  & 98.094 & \\
 3 & $\frac{3}{2}^-$ &  22.61     & 22.40(1) & 45.48(0) & 44.833    &  91.7(1)  & 90.915 & \\
 4 &  $0^+$          &  29.99     & 29.69(1) & 62.04(1) & 60.842    & 131.1(1)  & 126.31 & \\
 5 & $\frac{1}{2}^-$ &  41.22(1)  & 40.65(15)& 84.02(2) & 82.86(2)  & 175.2(1)  & 173.00 & \\
 5 & $\frac{3}{2}^-$ &  41.02     & 40.4(2)  & 82.97(1) & 80.68(2)  & 169.5(1)  & 162.71 & \\
 6 &  $0^+$          &  48.52(1)  & 47.6(2)  & 98.95(2) & 95.74(3)  & 205.8(2)  & 193.64  & $-$80.6\hphantom{(1)}  \\
 7 & $\frac{1}{2}^-$ &  59.17(1)  & 57.9(2)  & 118.9(0) & 115.67(5) & 246.4(2)  & 237.11  & $-$90.9(1) \\
 7 & $\frac{3}{2}^-$ &  59.73(1)  & 58.5(2)  & 121.1(0) & 118.9(1)  & 254.7(2)  & 249.85  & $-$88.6(1) \\
 8 &  $0^+$          &  67.01(1)  & 65.4(3)  & 135.8(0) & 132.5(1)  & 287.4(2)  & 278.32(1)&$-$103.9(1) \\
 9 & $\frac{1}{2}^+$ &  80.92(4)  & 78.9(1.5)& 163.7(1) & 159.6(4)  & 349.8(2)  & 334.32(1)&$-$107.8(1)\\
 9 & $\frac{3}{2}^+$ &            & 80.0(1.5)&          & 162.8(6)  & 354.5(2)  & 344.42(1)& \\
 9 & $\frac{5}{2}^+$ &  81.20(3)  & 79.3(1.5)& 163.2(1) & 159.4(4)  & 343.9(2)  & 331.15(1)&$-$106.6(1) \\
10 &  $0^+$         &   92.14(8)  & 90. (1.5)& 188.1(6) & 182.1(5)  & 400.5(2)  & 380.41(1)& $-$113.4(1) \\
11 & $\frac{1}{2}^+$ &  105.9(1)  &          & 216.1(3) & 208.4(1.0)&           & 434.38(5)& $-$115.9(2) \\
11 & $\frac{3}{2}^+$ &            &          &          & 208.0(1.0)&           & 430.10(4)\\
11 & $\frac{5}{2}^+$ &  106.3(1)  &          & 217.0(3) & 207.9(1.0)&           & 430.41(4)& $-$116.9(2) \\
12 &  $0^+$          &  118.1(1)  & 116.(6)  & 242.(1)  & 230.0(1.0)& 509.1(4)  & 477.05(5)& $-$123.6(3) \\
13 & $\frac{1}{2}^+$ &  130.8(1)  &          & 268.0(1.0)&255.8(1.0)&           & 529.07(6)& $-$125.0(3) \\
13 & $\frac{3}{2}^+$ &            &          &          & 256.3(1.0)&           & 528.74(6)\\
13 & $\frac{5}{2}^+$ &  131.5(1)  &          & 267.6(6) & 255.7(1.0)&           & 524.77(6)& $-$125.9(3) \\
14 &  $0^+$          &  142.2(2)  & 140.(10) & 291.9(2) & 277.5(1.4)&           & 569.3(1) & $-$131.6(7) \\
15 & $\frac{1}{2}^+$ &  160.1(1)  &          & 316.3(2) & 303.(5)   &           & 619.4(6) \\
15 & $\frac{3}{2}^+$ &  159.1(2)  &          & 320.2(2) &           &           &          \\
15 & $\frac{5}{2}^+$ &  160.0(1)  &          & 317.0(2) &           &           &          & $-$139.3(3) \\
16 & $0^+$           &  171.6(1)  &          & 341.5(2) & 326.(6)   & 730.3(3)  & 667.7(6) & $-$142.4(7) \\
17 & $\frac{1}{2}^+$ &  185.5(2)  &          & 368.8(3) &           &           & 725.1(8) & \\
17 & $\frac{3}{2}^+$ &  183.9(2)  &          & 366.5(3) & 352.(5)   &           &         & $-$148.8(2) \\
17 & $\frac{5}{2}^+$ &  184.9(2)  &          & 371.1(2) &           &           &         & \\
18 & $0^+$           &  195.6(2)  &          & 392.6(3) & 377.(7)   &           & 781.(1) & $-$155.1(4) \\
19 & $\frac{1}{2}^+$ &  209.4(2)  &          & 420.1(2) & 407.(7)   & 919.0(3)  & 850.(2) & \\
19 & $\frac{3}{2}^+$ &  208.4(2)  &          & 417.9(3) & 403.(7)   & 914.9(4)  & 838.(1) & $-$159.6(3) \\
19 & $\frac{5}{2}^+$ &  210.0(3)  &          & 422.1(2) & 408.(8)   & 926.7(4)  & 855.(2) & \\
20 & $0^+$           &  219.9(3)  &          & 441.7(4) & 430.(10)  & 976.0(4)  & 894.(2) & $-$165.0(1) \\
21 & $\frac{7}{2}^-$ &            &          & 476.8(4) & 465.(25)  &           & 956.(4) &  \\
22 & $0^+$           &  254.(1)   &          & 510.5(5) & 495.(25)  & 1123.3(7) & 1018.(5)&  \\
24 & $0^+$           &  289.1(7)  &          & 578.9(5) & < 596.    & 1268.(1)  & \le 1144. &  \\
26 & $0^+$           &  324.0(8)  &          & 645.0(6) & < 660.    &           & \le 1266. &  \\
28 & $0^+$           &  355.(1)   &          & 707.6(7) & < 723.    & 1551.(1)  & \le 1379. &  \\
30 & $0^+$           &  390.0(8)  &          & 776.0(9) & < 786.    &           & \le 1499. &  \\
32 & $0^+$           &  422.(1)   &          & 843.5(9) & < 847.    &           & \le 1614. &  \\
34 & $0^+$           &  453.(1)   &          & 909.9(9) & < 914.    &           & \le 1750. &  \\
36 & $0^+$           &  486.(1)   &          & 982.7(8) & < 986.    &           & \le 1895. &  \\
38 & $0^+$           &  514.(1)   &          &1046.4(8) & < 1057.   &           & \le 2037. &  \\
40 & $0^+$           &  546.(1)   &          &1114.3(9) & < 1128.   &           & \le 2177. &  \\
42 & $0^+$           &  591.(1)   &          &1197.1(8) & < 1206.   &           & \le 2320. &  \\
44 & $0^+$           &            &          &1278.(1)  &           &           & \le 2473. &  \\
%
\end{tabular}
\end{ruledtabular}
\caption{Total energy with AV8$^\prime$+UIX and JISP16 for several
  different confining wells.  Error bars for the AV8$^\prime$+UIX
  results are statistical only; in addition we expect systematic
  errors of a few percent, as discussed in Sec.~\ref{sec:afdmc_conv}.
  Error bars for the JISP16 results are total error estimates. (Errors
  that are not shown are less than 1 in the last digit.)
  \label{tab:totalE}}
\end{table*}
In Table~\ref{tab:totalE} we present the principal results of this
study: the total energies for neutron drops confined in 5 MeV, 10 MeV,
and 20 MeV HO wells with the AV8$^\prime$+UIX and JISP16 potentials.
These HO wells are convenient for ab-initio calculations because one
can probe very low to very high densities with a simple asymptotic
form of the wave function and an arbitrary number of neutrons can be
bound in the well.  In order to provide a very different probe of
density functionals in the extreme isospin limit, we also include
select results in a WS well with the AV8$^\prime$+UIX potential.

We show the lowest 0$^+$ energy for even $N$ and lowest values for
several $J^\pi$ for odd $N$.  We present results only for natural
parity states.  The AV8$^\prime$+UIX values up to $N$=14 were computed
by GFMC while the larger drops were computed using AFDMC, with the
exception of the 20 MeV HO well results.  There are no results from
GFMC available for the 20 MeV HO well, due to the strong fermion sign
problem with that external field strength; those results were all
obtained using AFDMC.

The JISP16 values were all computed by NCFC.  There are no results
from NCFC available in the 5 MeV trap above 14 neutrons due to poor
convergence with available computer resources.  Above 22 neutrons we
only provide strict upper bounds; for the 20 MeV HO well we expect the
converged energies to be within a few percent of these upper bounds.

\begin{figure}
\begin{center}
\center\includegraphics[width=0.99\columnwidth]{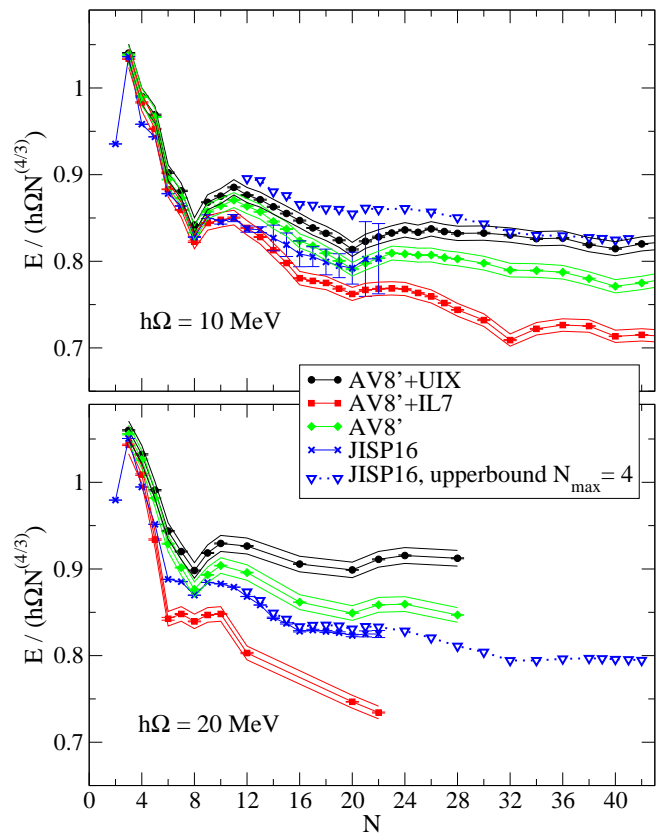}
\caption{(color online) 
  Energy of the lowest neutron drop states confined in a HO well with
  $\hbar\Omega=10$ MeV (top) and $\hbar\Omega=20$ MeV (bottom) as a
  function of the number of neutrons.  Results for AV8$^\prime$ (plus
  TNI) where obtained using AFDMC, with MC statistical error bars 
  as well as a band indicating the 1\% systematic uncertainty 
  discussed in Sec.~\ref{sec:afdmc_conv};
  results for JISP16 are obtained from NCFC with error bars reflecting
  the total numerical uncertainty, and strict upper bounds obtained
  with NCSM in finite basis spaces.  Note the pronounced dips at the
  expected HO magic numbers $N=2$, $8$, and $20$.
\label{fig:gs_scaled}}
\end{center}
\end{figure}
Figure~\ref{fig:gs_scaled} shows the energies of $N$ neutrons in two
different HO wells, scaled by $\hbar\Omega\,N^{(4/3)}$; for odd $N$
only the lowest energy found is shown.
The scaling by $\hbar\Omega\,N^{(4/3)}$ is motivated by the expected
results in local density approximation.  The factor $N^{4/3}$ comes
from the traditional scaling with $N$ times the increase in potential
energy arising from the increase in radius of the system with particle
number proportional to $N^{1/3}$.  In addition to the AV8$^\prime$+UIX
and JISP16 values presented in Table~\ref{tab:totalE}, we also show
results for AV8$^\prime$ without any TNI and AV8$^\prime$+IL7.  All
interactions show a very pronounced peak for three neutrons, and dips
at the expected HO magic numbers, $N=2$, $8$, $20$, and $40$.  The
dips at the HO magic numbers are expected due to the HO nature of the
confining well.

With an equation of state of the form 
$E = \xi \frac{\hbar^2}{2m} k_F^2$ with $k_F = [ 3 \pi^2 \rho ]^{1/3}$, 
the energy is given by the Thomas--Fermi expression: 
$E_{TF} = \xi^{1/2} \hbar\Omega\,(3N)^{(4/3)}/4$.  For free fermions 
($\xi =1$) the Thomas--Fermi results would be a horizontal line at
$3^{4/3}/4 \approx 1.081$, for a unitary Fermi gas with $\xi = 0.4$
the Thomas--Fermi results would be a horizontal line at 0.684.  
The calculated results are all below the free Fermi gas (even for the
case of three neutrons) since the interaction is attractive.  
All our results are above the unitary Fermi gas because 
there are significant finite-range corrections for neutron matter.
In addition repulsive gradient terms in the density functional are
required to reproduce the ab-initio results~\cite{Gandolfi:2010za}.  A detailed investigation
of these effects is being pursued.

From Fig.~\ref{fig:gs_scaled} it is evident that adding UIX to
AV8$^\prime$ increases the energies of neutron drops, whereas IL7
decreases the energies.  These results were expected; the two-pion
part of UIX is attractive in the isospin $T=1/2$ triples that appear
in nuclei.  However neutron drops contain only $T=3/2$ triples for
which the two-pion part is very small~\cite{Pudliner96,Gandolfi:2012};
this leaves only the repulsive central part of UIX.  On the other hand,
IL7 contains the three-pion term that is strongly attractive in
$T=3/2$ triplets~\cite{Pieper:2001}.

The energies with the nonlocal 2-body interaction JISP16
are generally below the AV8$^\prime$ results, but above the
AV8$^\prime$+IL7 ground state energies.  In fact, in the 10 MeV
HO trap, the JISP16 results are nearly identical to those with
AV8$^\prime$+IL7 up to about 12 neutrons; as the number of
neutrons increases, the results with JISP16 deviate more and more from
the AV8$^\prime$+IL7 results.  In the 20 MeV HO trap, for which
we have more accurate results with JISP16, the results with JISP16 and
with AV8$^\prime$ without TNI are quite similar, even in the
$sd$-shell and beyond.  The trend of the upper bounds obtained with
JISP16 follows the trend of the AV8$^\prime$ ground state energies
through the $sd$-shell and into the $pf$-shell, both in the 10 MeV and
in the 20 MeV trap.

As discussed in Sec.~\ref{sec:TNIs} and in
Sec.~\ref{sec:neutronmatter} below, recent studies of the neutron
star mass-radius relationship~\cite{Steiner:2012,Gandolfi:2012} suggest that,
at least at higher densities,  the AV8$^\prime$ +
UIX interactions gives a reasonable neutron matter equation of state.  
The requirement of a two-solar mass neutron star
implies a repulsive three-neutron interaction at moderate and high densities.

On the other hand, AV8$^\prime$+IL7 gives a much
better description of the ground state energies, spectra, and other
observables for light nuclei (up to $A=12$) than either AV8$^\prime$
or AV8$^\prime$+UIX. This may be why the results
with AV8$^\prime$+IL7 and with JISP16 (which also gives a good
description of light nuclei) are quite similar below 12 neutrons.
However, none of these interactions have been fit to any data beyond
the $p$-shell, and it is unclear which of these interactions is more
realistic for the neutron drops in the $N=8$ to $N=40$ range.
At larger densities AV8$^\prime$+IL7 is too attractive as discussed below.

In the 10 MeV well, the dips in the energies at $N=16$ and $N=32$
suggest subshell closure with AV8$^\prime$+IL7, but not with
AV8$^\prime$+UIX, while the results for AV8$^\prime$ show a hint of
subshell closure at $N=32$.  The IL7 TNI does provide a larger
spin-orbit splitting than the UIX three-nucleon interaction.
Similarly, the energies with JISP16
suggest subshell closure at $N=16$ and $N=32$ in the 20 MeV well.
The JISP16 results in the 10 MeV well are not quite accurate enough to
draw firm conclusion regarding subshell closure; and we have
insufficient results for AV8$^\prime$ in the 20 MeV well.

Somewhat surprisingly, there is no indication of subshell closure at
$N=28$.  In other words, these results seem to suggest closure of the
combined $f_\frac{7}{2}$ and $p_\frac{3}{2}$ subshell at $N=32$,
rather than closure of just the $f_\frac{7}{2}$ at $N=28$.  Note that
the closure of the combined $d_\frac{5}{2}$ and $s_\frac{1}{2}$
subshell at $N=16$ corresponds to the recently discovered subshell
closure at $^{24}$O~\cite{ClosedO24}.

\subsection{Energy differences}

\begin{figure}[thb]
\center\includegraphics[width=0.99\columnwidth]{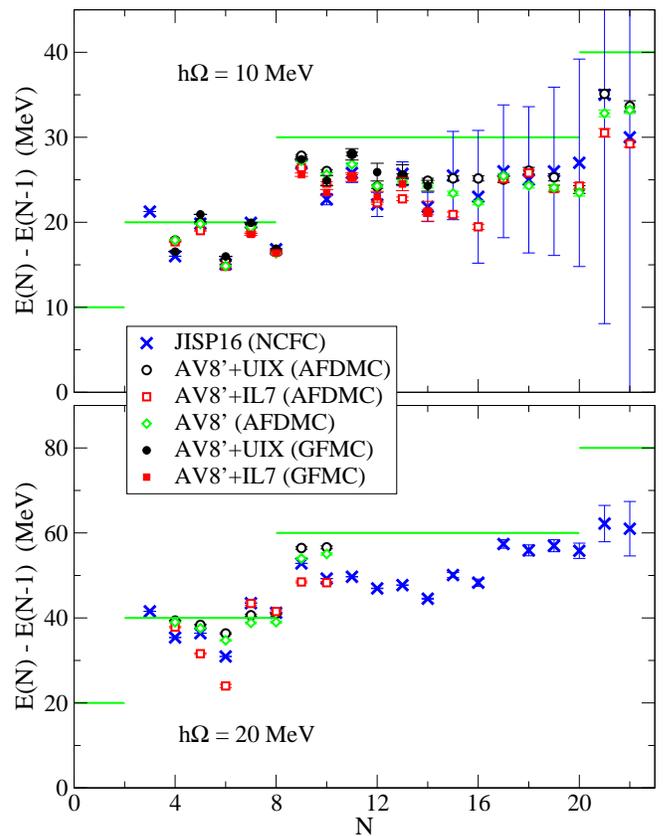}
\caption{(color online) Single energy differences in a 10 MeV (top)
  and 20 MeV (bottom) HO well.  Results of different Hamiltonians are
  compared.  AFDMC and GFMC error bars are statistical only; NCFC
  error bars reflect the total numerical uncertainty.  Horizontal line
  segments indicate energy differences expected from pure HO energies.
  \label{fig:sdiffs}}
\end{figure}
In Fig.~\ref{fig:sdiffs} we show the difference in total energy
between neutron drops with $N$ and with $N-1$ neutrons.  We clearly
see the effect of the HO shells: jumps at 2, 8, and 20 neutrons, at
which the next neutron has to go to the next HO shell.  Without
interactions between the neutrons, we would still have this shell
structure, but within each shell, all single energy differences would
be equal, as indicated by the solid reference lines in
Fig.~\ref{fig:sdiffs}.  That is, the gross feature of shell structure
arises from the confining well and is evident in the plot of the
single differences as a jump in the calculated energy differences as
one goes from one shell to the next.

\begin{figure}[thb]
\center\includegraphics[width=0.99\columnwidth]{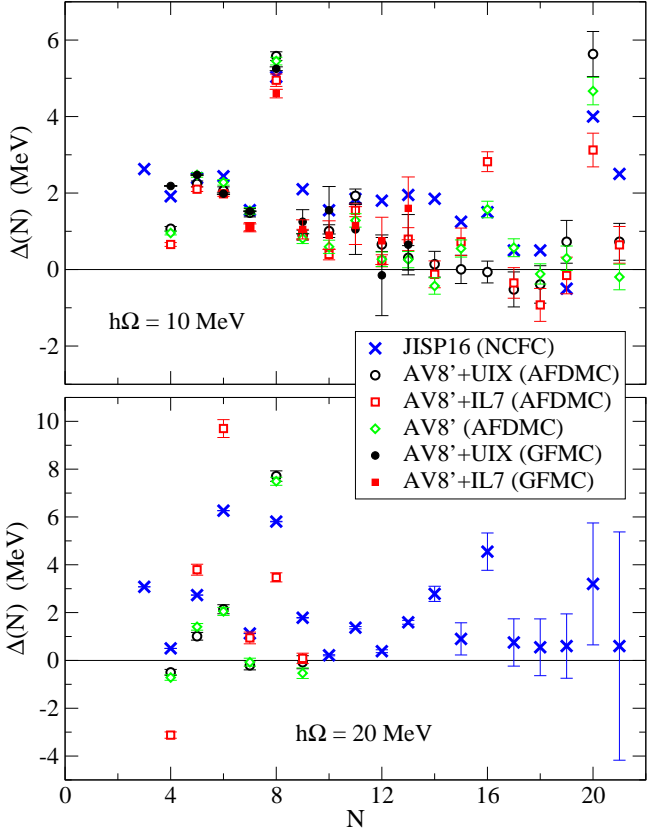}
\caption{(color online) 
  Double energy differences 
  $\Delta(N)=(-1)^{N+1} [E(N) - \frac{1}{2}\left(E(N-1) + E(N+1)\right)]$
  in a 10 MeV (top) and 20 MeV (bottom) HO well.  Results of different
  Hamiltonians are compared.  AFDMC and GFMC error bars are
  statistical only; the NCFC error bars are omitted for the 10 MeV HO
  well, because they would cover the entire vertical range for 12
  neutrons and above, though a significant part of the NCFC
  numerical error is systematic, and cancels between neighboring
  neutron drops; for completeness, we did include the total numerical
  uncertainty for the NCFC results in the 20 MeV HO well.  
  \label{fig:ddiffs}}
\end{figure}
The detailed fluctuations within a shell are entirely due to the
neutron interactions.  The most prominent feature is the neutron
pairing, in particular in the $p$-shell and also in the (beginning) of
the $sd$-shell.  This effect can be seen more clearly by looking at the
double difference in total energy 
$\Delta(N)=(-1)^{N+1} [E(N) - \frac{1}{2}\left(E(N-1) + E(N+1)\right)]$,
see Fig.~\ref{fig:ddiffs}.  The phase $(-1)^{N+1}$ is included to make
the pairing positive definite in the standard BCS theory.  Without
interactions, the double differences would be zero, except at the
magic numbers 2, 8, and 20.

Overall, the pairing seems to decrease as $N$ increases, except for
the closed (sub)shells.  Note that the pairing in nuclei also
decreases for larger nuclei~\cite{Bertsch:2012}.  On the other hand,
the numerical uncertainties increase with $N$, preventing us from
obtaining meaningful results for the pairing beyond 22 neutrons using
the NCFC approach.  AFDMC calculations of the pairing gaps will be
more reliable once BCS correlations have been included in the trial
state and this is being pursued.  For all methods we expect that the
error in neighboring neutron drops is correlated, resulting in a
reduced error in calculations of energy differences and pairing.

Despite the numerical uncertainties, there are some features that are
likely to be robust in Fig.~\ref{fig:ddiffs}.  As expected, the peaks
in the double difference $\Delta(N)$ at the magic numbers 8 and 20
stand out for all of the interactions for which we have results, in
particular for the 10 MeV HO well.  In addition, our results suggest
subshell closure at $N=16$ for AV8$^\prime$ without TNIs, with
AV8$^\prime$+IL7, and with JISP16, but not with AV8$^\prime$+UIX.
This closed subshell corresponds to the recently discovered subshell
closure at $^{24}$O~\cite{ClosedO24}, in which TNIs play a crucial
role.

In addition to the closure at $N=16$, we see evidence for subshell
closure at $N=6$ (the $p_\frac{3}{2}$) in the 20 MeV HO well both with
AV8$^\prime$+IL7 and with JISP16, but not in the 10 MeV HO well.  We
do not have sufficient data yet to examine the expected closure of the
$f_\frac{7}{2}$ at $N=28$, which was not evident in the plots of the
total energy (see Fig.~\ref{fig:gs_scaled}), nor for a more detailed
analysis of the closure at $N=32$ suggested in
Fig.~\ref{fig:gs_scaled}.

\subsection{Level splittings}

\begin{figure}[b]
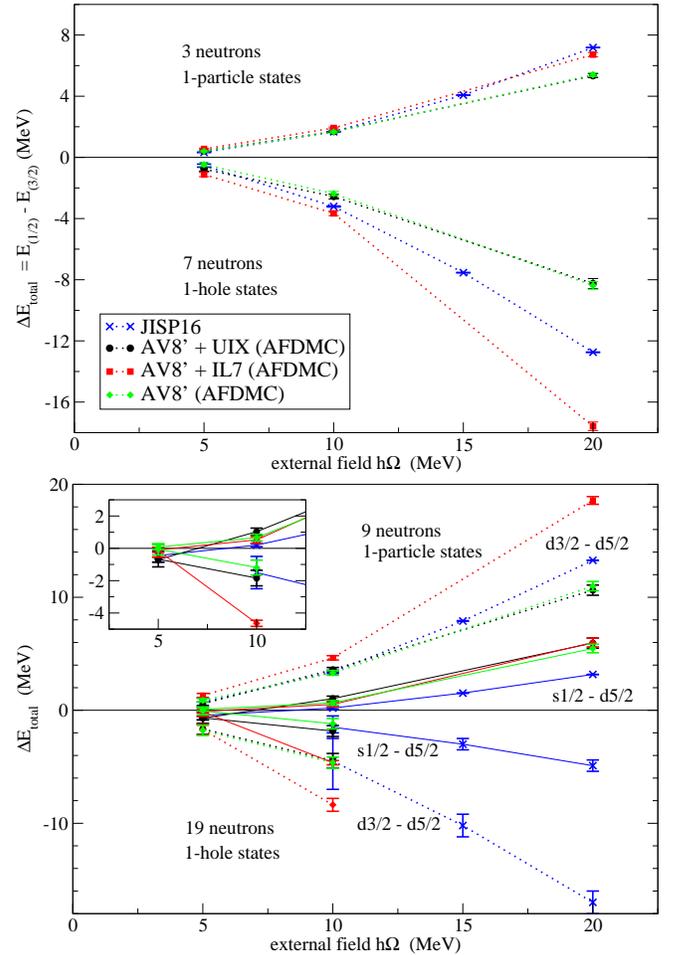

\begin{center}
\includegraphics[width=0.99\columnwidth]{splittings_p-shell}
\includegraphics[width=0.99\columnwidth]{splittings_sd-shell}
\caption{(color online) 
  Spin-orbit splitting in the $p$-shell (top)
  and level splittings in the $sd$-shell (bottom)
  and as a function of external field strength.
  Results of different Hamiltonians are compared.  
  Inset: blowup of the $s_\frac{1}{2}$ and $d_\frac{5}{2}$ levels
  for the 5 MeV and 10 MeV H.O. wells.
\label{fig:splittings}}
\end{center}
\end{figure}
If we look at the single-particle and single-hole states at the
beginning and end of the $p$-shell, see Fig.~\ref{fig:splittings},
we find that the spin-orbit splitting between the $\frac{1}{2}^-$ and
$\frac{3}{2}^-$ increases with $\hbar\Omega$ for all interactions.
For three neutrons, the splitting between these levels is almost the
same for JISP16 and AV8$^\prime$+IL7; however, for seven neutrons the
splitting is significantly enhanced with IL7.  On the other hand,
AV8$^\prime$ without TNI and AV8$^\prime$+UIX have almost the same
splitting.

The systematic increase in level splittings with increasing
$\hbar\Omega$ can be understood as follows: With increased
$\hbar\Omega$, the radial shape is increasingly constrained by the HO
potential and the associated gaussian falloff of the radial densities
in the surface region.  This increase in level splittings with
$\hbar\Omega$ may then be interpreted as a consequence of the
increased density gradient in the surface region.

In the $sd$-shell the splitting between the $d_\frac{5}{2}$ and
$s_\frac{1}{2}$ levels (solid lines in Fig.~\ref{fig:splittings}) is
much smaller than the splitting between these two levels and the
$d_\frac{3}{2}$ level, in particular for AV8$^\prime$+IL7.  This
confirms the subshell closure at $N=16$ that was evident from the
pairing, see Fig.~\ref{fig:ddiffs}.  It is also in apparent agreement
with the observation in known nuclei that the subshell closure at 16
neutrons (both $d_\frac{5}{2}$ and $s_\frac{1}{2}$ levels filled) is
much stronger than the subshell closure at 14 neutrons (only the
$d_\frac{5}{2}$ level filled).

Furthermore, notice that the level ordering can change as the strength
of the HO well increases in the case of 9 neutrons: in the weakest
well of 5 MeV (i.e. at very low density), the $s_\frac{1}{2}$ is
slightly below the $d_\frac{5}{2}$ level, but as $\hbar\Omega$
increases, the $d_\frac{5}{2}$ becomes the lowest level.
Interestingly, this happens both with JISP16 and with
AV8$^\prime$+IL7, whereas with AV8$^\prime$+UIX and with AV8$^\prime$
(without TNIs) the $d_\frac{5}{2}$ and $s_\frac{1}{2}$ are basically
degenerate for the 5 MeV HO well.

In the $pf$-shell we find qualitatively similar results with JISP16: a
large spin-orbit splitting between the $f_\frac{7}{2}$ and
$f_\frac{5}{2}$ levels and between the $p_\frac{3}{2}$ and
$p_\frac{1}{2}$ levels, a smaller splitting between the
$p_\frac{1}{2}$ and $f_\frac{5}{2}$ levels, and an even smaller
splitting between the $f_\frac{7}{2}$ and $p_\frac{3}{2}$ levels.  All
of these level splittings increase significantly with the strength of
the HO well: at $\hbar\Omega = 5$~MeV, the splittings are almost
negligible, less than an MeV, and within the numerical uncertainty.
On the other hand, at $\hbar\Omega = 20$~MeV (the largest value that
we have considered) the spin-orbit splittings are of the order of
ten(s) of MeV.

\subsection{Internal energies and radii}

\begin{table*}
\renewcommand{\arraystretch}{1.2}
\begin{ruledtabular}
\begin{tabular}{rr|dd|dd|dd|dd|dd|dd}
     &        & \multicolumn{4}{c|}{ 5 MeV HO well} 
              & \multicolumn{4}{c|}{10 MeV HO well} 
              & \multicolumn{2}{c|}{20 MeV HO well} 
              & \multicolumn{2}{c}{WS well}  \\ 
\hline
 $N$ & $J^\pi$ & \multicolumn{2}{c|}{AV8$^\prime$+UIX} & \multicolumn{2}{c|}{JISP16}
              & \multicolumn{2}{c|}{AV8$^\prime$+UIX} & \multicolumn{2}{c|}{JISP16}
              & \multicolumn{2}{c|}{JISP16}         & \multicolumn{2}{c}{AV8$^\prime$+UIX} \\
\hline
 3 & $\frac{1}{2}^-$ &  11.5 & 3.56 &  11.2    & 3.58    &  22.3 & 2.60 &  22.02  & 2.60   &  44.31 & 1.928(1)\\
 3 & $\frac{3}{2}^-$ &  11.3 & 3.52 &  11.2    & 3.51    &  22.3 & 2.53 &  22.26  & 2.50   &  43.80 & 1.805(1)\\
 4 &  $0^+$          &  14.8 & 3.55 &  14.4    & 3.56    &  29.1 & 2.61 &  29.10  & 2.57   &  58.92 & 1.869(1)\\
 5 & $\frac{1}{2}^-$ &  20.6 & 3.70 &  20.1(2) & 3.70(2) &  40.1 & 2.70 &  39.8   & 2.67   &  79.5  & 1.969(2)\\
 5 & $\frac{3}{2}^-$ &  20.6 & 3.68 &  20.3(2) & 3.65(2) &  40.0 & 2.70 &  40.5   & 2.58   &  78.5  & 1.869(2)\\
 6 &  $0^+$          &  24.2 & 3.67 &  23.8(4) & 3.63(3) &  47.3 & 2.67 &  47.8   & 2.58   &  92.8 & 1.867(1)& 46.9 &   2.70 \\
 7 & $\frac{1}{2}^-$ &  29.7 & 3.74 &  29.7(5) & 3.66(3) &  56.8 & 2.71 &  57.3(2)& 2.63   & 111.3(1)& 1.930(2)& 55.6 &   2.75 \\
 7 & $\frac{3}{2}^-$ &  29.9 & 3.76 &  29.7(5) & 3.70(3) &  57.2 & 2.75 &  56.9(2)& 2.71   & 113.2(1)& 2.012(2)& 55.2 &   2.81 \\
 8 &  $0^+$          &  35.0 & 3.64 &  33.2(6) & 3.65(3) &  64.4 & 2.72 &  63.5(2)& 2.68(1)& 126.2(1)& 1.986(2)& 63.2 &   2.75 \\
 9 & $\frac{1}{2}^+$ &  41.9 & 3.79 &  40.(2)  & 3.77(5) &  77.9 & 2.81 &  76.5(1.)& 2.77(2)& 152.7(2)& 2.045(2)& 70.5 &   3.03 \\
 9 & $\frac{3}{2}^+$ &       &      &  41.(2)  & 3.82(5) &       &      &  77.1(1.)& 2.81(3)& 155.2(2)& 2.088(2)&\\
 9 & $\frac{5}{2}^+$ &  42.2 & 3.79 &  41.(2)  & 3.78(5) &  78.3 & 2.80 &  77.4(1.)& 2.75(2)& 153.2(2)& 2.024(2)& 73.6 &   2.93 \\
10 &  $0^+$          &  46.7 & 3.88 &  45.(2)  & 3.85(10)&  88.  & 2.89 &  87.0(1.)& 2.81(2)& 176.0(3)& 2.059(2)& 78.  &   3.11 \\
11 & $\frac{1}{2}^+$ &  53.7 & 3.97 &          &         &  99.  & 2.97 & 100.(2.) & 2.86(3)& 201.7(8)& 2.095(4)& 89.  &   3.21 \\
11 & $\frac{3}{2}^+$ &       &      &          &         &       &      & 101.(2.) & 2.84(3)& 202.1(5)& 2.074(3)&\\
11 & $\frac{5}{2}^+$ &  53.2 & 4.00 &          &         & 102.  & 2.95 & 100.(2.) & 2.85(3)& 202.1(5)& 2.075(3)& 87.  &   3.23 \\
12 &  $0^+$          &  59.4 & 4.03 &          &         & 110.  & 3.02 & 110.(2)  & 2.88(3)& 224.0(5)& 2.090(3)& 97.  &   3.20 \\
13 & $\frac{1}{2}^+$ &  65.5 & 4.08 &          &         & 121.  & 3.06 & 123.(2)  & 2.91(3)& 248.5(6)& 2.116(3)&106.  &   3.29 \\
13 & $\frac{3}{2}^+$ &       &      &          &         &       &      & 123.(2)  & 2.91(3)& 249.5(6)& 2.111(3)& \\
13 & $\frac{5}{2}^+$ &  65.9 & 4.09 &          &         & 120.  & 3.06 & 124.(2)  & 2.90(3)& 249.9(6)& 2.094(3)&103.  &   3.37 \\
14 &  $0^+$          &  71.1 & 4.10 &          &         & 132.  & 3.08 & 134.(3)  & 2.92(4)& 271.8(7)& 2.099(3)&115.  &   3.31 \\
\end{tabular}
\end{ruledtabular}
\caption{
  Internal energies (in MeV) and rms radii (in fm) of neutron drops
  with various external potentials and a selected set of Hamiltonians.
  The results are plotted in Figs.~\ref{fig:int} and \ref{fig:radii}.
  The particular many-body method used to produce these results
  depends on the external field, the number of neutrons and the
  Hamiltonian as described in the text.
  \label{tab:intErad}}
\end{table*}
In Table~\ref{tab:intErad} we list our results for the internal energy
$E_{\hbox{\scriptsize int}} = 
\langle H \rangle - \langle U_{\hbox{\scriptsize ext}} \rangle$,
as well as for the rms radii for systems up to 14 neutrons
in a HO well with JISP16 and with AV8$^\prime$+UIX,
as well as in a WS well with AV8$^\prime$+UIX.  Note that
for neutron drops in a HO well the radius is directly related to the
external energy, $\langle U_{\hbox{\scriptsize ext}} \rangle = 
\frac{1}{2} m \omega^2 \langle r^2 \rangle$
for a HO external field.  Overall, the internal energy is typically
slightly less than half of the total energy (see
Table~\ref{tab:totalE} for comparison), 
and $\langle U_{\hbox{\scriptsize ext}} \rangle$ is slightly more than
half the total energy.  This is to be expected, since the total energy
scales approximately as $\hbar\Omega\,N^{(4/3)} \propto \rho^{2/3}$,
and for all cases where the equation of state is proportional to
$\rho^{2/3}$, the virial theorem will give equal internal energies and
one-body potential energies, each one-half of the total energy.

\begin{figure}
\center\includegraphics[width=0.99\columnwidth]{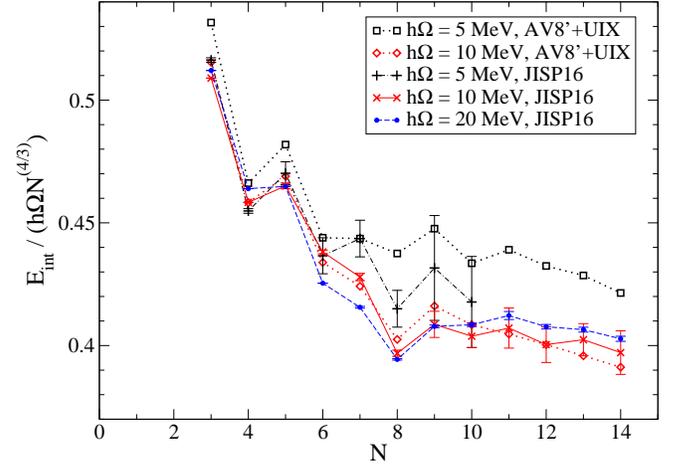}
\caption{(color online) 
  Internal energy for up to 14 neutrons in a HO trap with
  AV8$^\prime$+UIX and with JISP16.  For details see
  Table~\ref{tab:intErad}.
  \label{fig:int}}
\end{figure}
\begin{figure}
\begin{center}
\includegraphics[width=0.99\columnwidth]{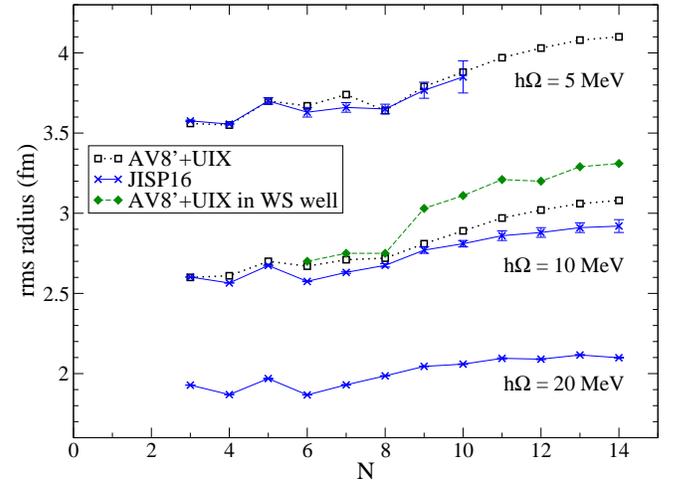}
\caption{(color online) 
  Radii for the lowest $J$ states up to 14 neutrons in a HO trap with
  AV8$^\prime$+UIX and with JISP16.  For details see
  Table~\ref{tab:intErad}.
  \label{fig:radii}}
\end{center}
\end{figure}

In Fig.~\ref{fig:int} we show the internal energy
$E_{\hbox{\scriptsize int}}$, scaled by $\hbar\Omega\,N^{(4/3)}$, of
the lowest $J=0$ and $J=\frac{1}{2}$ states for up to 14 neutrons in a
HO well.  In the 10 MeV trap the JISP16 and the AV8$^\prime$+UIX
results are rather close to each other, significantly closer than
the total energies shown in Fig.~\ref{fig:gs_scaled}.  Apparently, the
larger differences observed in Fig.~\ref{fig:gs_scaled} arise
primarily from differences in their 
$\langle U_{\hbox{\scriptsize ext}} \rangle$ energy shifts.  Indeed,
the corresponding rms radii, 
and thus $\langle U_{\hbox{\scriptsize ext}} \rangle$, start to deviate
from each other above $N=10$, see Fig.~\ref{fig:radii}.
The two interactions also give quite similar internal energy results
in the 5 MeV trap as seen in Fig.~\ref{fig:int}, given the rather
large error bars of the NCFC results, and the corresponding radii are
almost identical, at least up to 10 neutrons.

Table~\ref{tab:intErad} shows that both the internal energies and the
rms radii in the WS well are of the same order as those in the 10 MeV HO
trap, even though the total energies are very different.  In fact, the
rms radii are nearly identical for the three $p$-shell neutron drops
($N=6$, $7$, and $8$), but in the $sd$-shell there are significant
differences between the HO and the WS radii.

We note that the internal energy Fig.~\ref{fig:int} displays a similar
odd-even effect due to pairing as we observed in the total energy in
Fig.~\ref{fig:gs_scaled}.  The radii of the $J=0$ and $J=\frac{1}{2}$
states also show an odd-even effect, but only in the $p$-shell, for
three to seven neutrons; there is no a significant odd-even effect for
these states above $N=8$ in Fig.~\ref{fig:radii}.  Also note that the
radii of states with different $J$ in odd neutron systems are slightly
different, in particular in the $p$-shell, and more so with JISP16
than with AV8$^\prime$ + UIX, as can be seen in
Table~\ref{tab:intErad}.

\subsection{Densities}

\begin{figure}
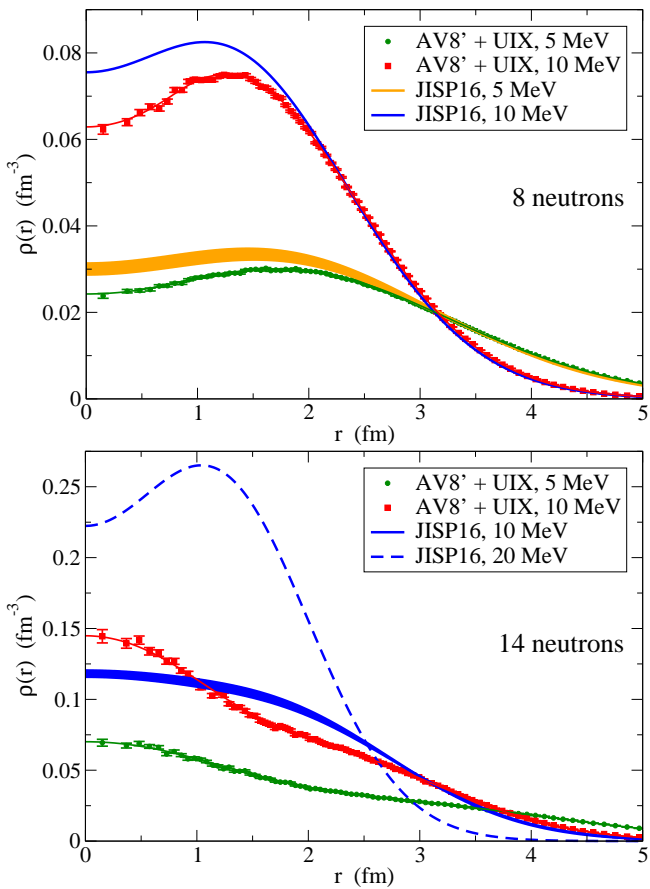

\begin{center}
\includegraphics[width=0.99\columnwidth]{density_Combined_N8}
\includegraphics[width=0.99\columnwidth]{density_Combined_N14}
\caption{(color online) 
  Radial density distributions for 8 (top) and 14 (bottom) neutrons in
  different HO traps with JISP16 and with AV8$^\prime$+UIX.
  \label{densities}}
\end{center}
\end{figure}
We present a sample set of radial density distributions computed with
JISP16 and with the AV8$^\prime$+UIX Hamiltonian in
Fig.~\ref{densities}.  The band thickness of the JISP16 results,
obtained with NCFC, is our best estimate of the total numerical
uncertainty in these densities; the AV8$^\prime$+UIX results were
calculated with GFMC, and the error bars correspond to the statistical
errors in the GFMC approach.  Given the HO nature of the trap, all
density distributions fall like gaussians at distances sufficiently
far from the origin.

The various densities for 8 neutrons (top panel of
Fig.~\ref{densities}, closed $p$-shell) are quite similar for the two
different interactions.  The only difference is that the central
densities, below 1 fm, are about 10\% to 20\% higher with JISP16 than
with AV8$^\prime$+UIX, but the shape is essentially the same, and
above 2 to 3~fm the densities are practically on top of each other.
This could be expected from the similar rms radii for these cases
shown in Table~\ref{tab:intErad}.  As the HO trap strength increases,
the density distribution gets compressed, the rms radius decreases,
and the central density increases, as one would expect.  The radial
shape, a slight dip at $r=0$, is typical for the closed $p$-shell, and
is qualitatively the same for weak and strong HO traps.

However, the shape of the density profile for 14 neutrons is somewhat
different for the two interactions; furthermore, the shape seems to
depend on the strength of the HO trap, at least for JISP16.  With
JISP16 in the 20 MeV trap (dashed curve in the bottom panel of
Fig.~\ref{densities}), the density has a clear dip at the center, and
peaks at a distance of about 1~fm from the center.  In fact, the shape
of this density is rather similar to that of 8 neutrons in a HO trap.
On the other hand, in the 10 MeV trap there is no evidence for such a
dip; rather, within the estimated numerical accuracy, the density
seems to fall off monotonically from a central value of about 0.11 to
0.12 fm$^{-3}$ with JISP16.  On the other hand, the densities obtained
with the AV8$^\prime$+UIX Hamiltonian for 14 neutrons seem to be
slightly enhanced in the central region: in the 10 MeV trap the
central density with with the AV8$^\prime$+UIX is about 20\% higher
than with JISP16.

This difference between the density profiles of 14 neutrons in a HO
trap with JISP16 and AV8$^\prime$+UIX could well be related to the
presence (JISP16) and absence (AV8$^\prime$+UIX) of sub-shell closure
for 16 neutrons; and both are likely to be related to differences in
spin-orbit splittings.  It would be interesting to compare these
densities with those obtained with other realistic potentials, and in
particular to investigate the effect of different 3-body forces on the
density profiles as well as on the spin-orbit splittings and sub-shell
closures.

\section{Neutron matter
\label{sec:neutronmatter}}

\begin{figure}
\begin{center}
\includegraphics[width=0.4\textwidth]{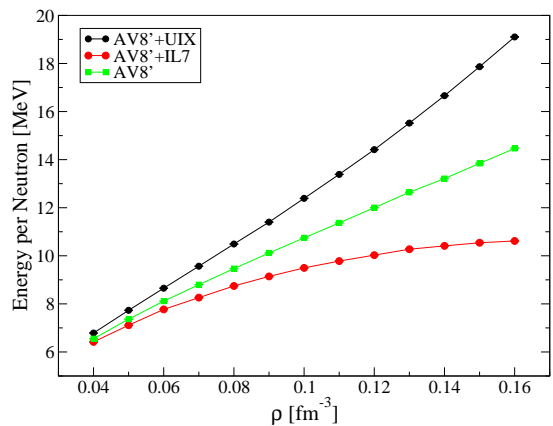}
\caption{(color online) 
  Equation of state of neutron matter as a function of the density for
  different Hamiltonians.  
  \label{fig:eos}}
\end{center}
\end{figure}

The equation of state of neutron matter is important to properly fix
the bulk term of Skyrme-type EDFs.  We report the AFDMC results for
the energy per neutron as a function of the density in
Table~\ref{tab:eos} and display them in Fig.~\ref{fig:eos}.  For the
AFDMC Quantum Monte Carlo calculations, the results are for a system
of 66 neutrons with periodic boundary conditions.  The calculation is
very similar to those of the neutron drops, except the single-particle
orbitals in the trial wave function are replaced by plane waves that
respect the periodic boundary condition, as described in
Sec.~\ref{sec:afdmc_wf}.  More details can be found in
Refs.~\cite{Gandolfi:2009,Sarsa:2003}.  The energy corrections due to
finite-size effects arising from such a simulation are expected to be
extremely small compared to the bulk energies considered here.

\begin{table}
\begin{center}
\begin{tabular}{c|ccc}
\hline
$\rho$ [fm$^{-3}$] & AV8$^\prime$ & AV8$^\prime$+UIX & AV8$^\prime$+IL7 \\
\hline
0.04               &  6.55(1)     &  6.79(1)         &  6.42(1) \\
0.05               &  7.36(1)     &  7.73(1)         &  7.11(1) \\
0.06               &  8.11(1)     &  8.65(1)         &  7.77(1) \\
0.07               &  8.80(1)     &  9.57(1)         &  8.26(1) \\
0.08               &  9.47(1)     & 10.49(1)         &  8.75(2) \\
0.09               & 10.12(1)     & 11.40(1)         &  9.14(2) \\
0.10               & 10.75(1)     & 12.39(1)         &  9.50(2) \\
0.11               & 11.37(1)     & 13.39(1)         &  9.78(2) \\
0.12               & 12.00(1)     & 14.42(1)         & 10.03(2) \\
0.13               & 12.64(1)     & 15.52(1)         & 10.27(2) \\
0.14               & 13.21(1)     & 16.66(1)         & 10.41(2) \\
0.15               & 13.84(2)     & 17.87(2)         & 10.54(3) \\
0.16               & 14.47(2)     & 19.10(2)         & 10.62(3) \\
\hline
\end{tabular}
\caption{Equation of state of neutron matter as a function of the
  density for various Hamiltonians.  
\label{tab:eos}}
\end{center}
\end{table} 

The effect of TNI is important in the equation of state of neutron
matter beyond half nuclear matter saturation density ($\rho =
0.08~{\rm fm}^{-3}$), as is clear in Fig.~\ref{fig:eos}.  The two
different TNIs added to AV8$^\prime$ have opposite effects: UIX is
repulsive, while IL7 is attractive.  This is in agreement with the
trend in neutron drops shown in Fig.~\ref{fig:gs_scaled}.  Our earlier
discussion of the effects of the different terms in UIX and IL7 apply
equally to the differences observed here.  Furthermore, in moderately
large neutron drops ($N > 12$) we have seen that the trend with JISP16
is similar to the trend with AV8$^\prime$ without TNI.  We therefore
expect that the equation of state with JISP16 will be similar to that
of AV8$^\prime$ without TNIs.

The equation of state of pure neutron matter with the AV8$^\prime$ NN
interaction alone is rather soft at high densities.  It is only
marginally compatible with the recently-observed two-solar mass
neutron
star~\cite{Demorest:2010,Gandolfi:2012,Akmal:1998,Steiner:2012}.  The
relevant three-neutron force must be, in aggregate, repulsive at high
densities~\cite{Akmal:1998}.  While the three-pion terms in IL7 (and
in $\chi$PT forces) are certainly present, they are far too attractive
in the IL7 model alone.  It is possible to adjust the short- and
long-range three neutron forces to be repulsive by varying the
contributions of these magnitudes, as obtained in the recent studies
of three-neutron interactions and the neutron star mass-radius
relations~\cite{Gandolfi:2012}.

\section{Conclusions
\label{sec:conclusion}}

We have computed the properties of neutron drops confined by external
harmonic oscillator (HO) and Woods Saxon (WS) traps using a variety of
realistic nucleon-nucleon and nucleon-nucleon plus three-nucleon
interactions (TNIs).  The combination of results with HO and WS wells
should prove useful in separating bulk and gradient (surface) effects,
and testing the general form of the density functional.  The pairing
and spin orbit splittings may also have very different behavior.

We have employed currently available state-of-the art many-body
methods to obtain these results and we have quantified the
uncertainties in the results.  We observe characteristic features such
as pairing and subshell closures in qualitative agreement with
expectations.  Differences in results for the same systems are
attributable in large part to differences in the interactions.  We
present and interpret significant sensitivity of some obervables to
the TNI. The radii of the neutron drops appear to be rather robust -
that is approximately independent among the interactions we employ.

The results we obtain for the neutron equation of state as a function
of density follow trends seen in the neutron drop results as a
function of the external HO trap.  This is significant since the
neutron drops have quantified uncertainties on their total and
internal energies as well as their rms radii, while it is more
difficult to quantify the uncertainty in the calculated neutron matter
equation of state.

We anticipate that results for these extreme and idealized systems may
serve as guides to experiments on very neutron-rich nuclei.  We also
hope these results will inform developments of improved energy-density
functionals~\cite{Gandolfi:2010za,Kortelainen:2011ft}.

\begin{acknowledgments}
We thank G. F. Bertsch, S. Bogner, A. Bulgac, F. Coester,
J. Dobaczewski, W. Nazarewicz, S. Reddy, A. Shirokov and R. B. Wiringa
for valuable discussions.
This work is supported by the U.S. DOE SciDAC program 
through the NUCLEI collaboration, 
by the U.S. DOE Grants
DE-SC-0008485 (SciDAC/NUCLEI), 
DE-FG02-87ER40371, 
and by the U.S. DOE Office of Nuclear Physics under Contracts 
DE-AC02-06CH11357, 
and DE-AC52-06NA25396. 
This work is also supported by 
the U.S. NSF Grant 0904782, 
and by the LANL LDRD program.  
We thank the Institute for Nuclear Theory at the University of 
Washington for its hospitality and the DOE for partial support 
during various stages of this work.
Computer time was made available by Argonne's LCRC, 
the Argonne Mathematics and Computer Science Division, 
Los Alamos Institutional Computing, 
the National Energy Research Scientific Computing Center (NERSC),
which is supported by the DOE Office of Science under Contract
DE-AC02-05CH11231,
and by an INCITE award, Nuclear Structure and Nuclear Reactions, 
from the DOE Office of Advanced Scientific Computing.  
This research used resources of the Oak Ridge Leadership Computing
Facility at ORNL, which is supported by the DOE Office of Science
under Contract DE-AC05-00OR22725,
and of the Argonne Leadership Computing Facility at ANL, which is
supported by the DOE Office of Science under Contract
DE-AC02-06CH11357.
\end{acknowledgments}

\bibliography{biblio}

\end{document}